\documentclass[11pt, a4paper]{article}
\pdfoutput=1

\usepackage[T1]{fontenc}
\usepackage[latin1]{inputenc}
\usepackage{lmodern}
\usepackage{amsmath}
\usepackage[pdftex]{graphicx}
\usepackage{hyperref}

\usepackage{braket} 
\usepackage{amssymb} 

\usepackage{titling}
\usepackage{authblk}

\usepackage{esvect}
\renewcommand*\vec{\vv}

\newcommand*\diff{\mathrm{d}} 
\let\limitint\int  \renewcommand{\int}{\limitint \!} 
\newcommand*{\disp}{ \omega_{\vec{k}}} 
\newcommand*\intMode{\int \frac{\diff^3 \vec{k}}{\sqrt{(2\pi)^3 2\disp}}}

\title{Quantum Break-Time of de Sitter\\[1.5\baselineskip]}

\author[1,2,3]{Gia Dvali}
\author[1, 4]{C\'{e}sar G\'{o}mez}
\author[1,2]{Sebastian Zell\thanks{sebastian.zell@campus.lmu.de}}
\affil[1]{Arnold Sommerfeld Center, Ludwig-Maximilians-Universit\"at, \mbox{Theresienstraße 37, 80333 M\"unchen, Germany}}
\affil[2]{Max-Planck-Institut für Physik, F\"ohringer Ring 6, 80805 M\"unchen, Germany}
\affil[3]{Center for Cosmology and Particle Physics, Department of Physics,\mbox{ New York University, 4 Washington Place, New York, NY 10003, USA}}
\affil[4]{Instituto de F\'{i}sica Te\'orica UAM-CSIC, Universidad Aut\'onoma de Madrid, Cantoblanco, 28049 Madrid, Spain}

\begin{document}

\maketitle

\begin{abstract}
 The quantum break-time of a system is the time-scale after which its true quantum evolution departs from the classical mean field evolution.  For capturing it, a quantum resolution of the classical background -- e.g., in terms of a coherent state -- is required. In this paper, we first consider a simple scalar model with anharmonic oscillations and derive its quantum break-time. Next, following \cite{Dvali:2013Compositness}, we apply these ideas to de Sitter space.  We formulate a simple model of a spin-2 field, which for some time reproduces the de Sitter metric and simultaneously allows for its well-defined representation as quantum coherent state of gravitons.  The mean occupation number $N$ of background gravitons turns out to be equal to the de Sitter horizon area in Planck units, while their frequency is given by the de Sitter Hubble parameter.  In the semi-classical limit, we show that the model reproduces all the known properties of de Sitter, such as the redshift of probe particles and thermal Gibbons-Hawking radiation, all in the language of quantum $S$-matrix scatterings and decays of coherent state gravitons. Most importantly, this framework allows to capture the $1/N$-effects to which the usual semi-classical treatment is blind. They violate the de Sitter symmetry and lead to a finite quantum break-time of the de Sitter state equal to the de Sitter radius times $N$. We also point out that the quantum-break time is inversely proportional to the number of particle species in the theory. Thus, the quantum break-time imposes the following consistency condition: Older and species-richer universes must have smaller cosmological constants.  For the maximal, phenomenologically acceptable number of species, the observed cosmological constant would saturate this bound if our Universe were $10^{100}$ years old in its entire classical history. 
\end{abstract}
\clearpage
\vspace*{-0.13\textheight}

\tableofcontents
\clearpage

\section{Introduction of the basic concept}
\label{sec:Introduction}

\subsection{The goal}

\subsubsection*{The world is fundamentally quantum}
  At the fundamental level, nature is quantum, i.e., 
  Planck's constant $\hbar$ is finite. Despite that fact, the classical approximation 
  works very well in many cases.  Namely, this is true for  systems with large occupation 
  number of quanta, $N \gg 1$. They behave more and more classically with 
  $N \rightarrow \infty$.  The precise meaning of the last statement is the following.  
  In most of such cases the notion of a {\it mean field}  can be defined by replacing the creation and annihilation operators of quanta by their $c$-number expectation values.  The dynamics of the system then can be 
  described by the {\it classical} evolution of  this mean field which obeys 
 certain classical equations of motion.  Usually, this classical description in terms of a given mean field is good up to $1/N$-effects, i.e., up to effects 
which allow to resolve a given state in its quantum constituents. 

 For example, in quantum electrodynamics a coherent state of photons of a given frequency and a large mean occupation number is well described by a classical electromagnetic wave. The role of the mean field in this case is played by a classical electromagnetic field, which obeys classical Maxwell's equations.  This description is excellent up to observations which are sensitive to individual photons.  Obviously, such  
effects correct the evolution of the system by $1/N$ relative to the leading order classical evolution. 
  
  \subsubsection*{The quantum break-time}
  How long does the classical description in terms of a particular mean field last?   In order to answer this question, one needs to introduce another quantum characteristic of the system: the quantum coupling $\alpha$. Usually, we can define this parameter as a dimensionless quantity which measures the characteristic strength of the two-particle scattering amplitude.  We shall only be interested in theories (or domains of applicability of a given theory) for which the quantum coupling is weak, i.e.,  $\alpha \ll 1$. By its very definition,  $\alpha$ must vanish for $\hbar = 0$. 
 
  Consider now a system in a quantum state with large $N$ which at some initial time  $t=0$ is well-described by the classical dynamics of some mean field  $\Phi$.  Usually, we can define a {\it collective coupling}  $\alpha N$, which is independent of $\hbar$ and thus represents a classical quantity. This collective coupling can be  viewed as the strength of classical nonlinear interactions of the mean field $\Phi$.  Correspondingly, although such classical nonlinearities can be large, they do not jeopardize the classical description of the system.  Notice that $\alpha$ is a parameter of the theory, whereas $N$ is a parameter of a given quantum state and so is the collective coupling $\alpha N$.  So,  for given $\alpha$, the  collective coupling  can be weak or strong,  depending on 
  $N$.  If in a given state the collective coupling is weak, $\alpha N \ll 1$, the classical mean field solution can be 
found in a perturbative series in powers of  $\alpha N$. 
  
 Inevitably,  since the system is interacting, there exist quantum processes  to which this mean field description is blind. These are processes which take into account rescatterings of the individual quanta  and are suppressed by powers of  $\alpha$ relative to the classical nonlinearities. These processes are important because they lead to departures of the true quantum evolution of the system from the classical mean field description 
and cause a complete breakdown of the latter description after some time. Following \cite{Nico}, we shall refer to this time-scale as {\it quantum break-time}  $t_{\text{q}}$. From the above argument it follows that essentially any classical system has an associated quantum break-time. 
  
  \subsubsection*{Classical Gravity as mean field effect}
  Although it seems completely generic, the above concept of understanding classical solutions as mean field description of an underlying quantum state was not applied to gravitational systems for a long time. In part, this is not surprising since the quantum resolution of nonlinear systems is not an easy task, due to 
 difficulties in defining correct creation/annihilation operators. In gravitational systems, such as Schwarzschild or de Sitter space-times, an additional difficulty arises from the existence of horizon, which makes it impossible to globally define the notion of  time.
 
  Only recently, first steps towards a formulation of a quantum description of classical gravitational systems were taken. As a first application, it was suggested that black holes can be understood as multi-particle states 
of $N$ soft gravitons \cite{Dvali:2011NPortrait}. It was shown that such a resolution of classical gravitational field implies that the occupation number $N$ numerically coincides with the black hole entropy. At the same time, the individual graviton-graviton coupling comes out to be inversely related to it:  $\alpha = 1/N$. Thus, despite the fact that for large $N$ the graviton-graviton coupling is minuscule, the collective
coupling is always strong:
\begin{equation}
\alpha N =1\, .
\label{critical} 
\end{equation} 
As it turns out, this relation has a deep physical meaning. In particular, it was shown that a in many-body language, equation \eqref{critical}  means that the $N$-graviton state describing a black hole is near a certain {\it quantum critical point} \cite{Dvali:2012:PhaseTransition}. 
This criticality is what provides the nearly-gapless {\it collective}  modes, which are necessary for accounting for the entropy of a black hole.  These quantum modes are related with the graviton constituents by a Bogoliubov transformation and have energy gaps suppressed by $1/N$. These modes are not detectable in the semi-classical treatment, which corresponds to taking the limit  $N=\infty$. Thus, the  finite-$N$ coherent state resolution of a black hole
is crucial for uncovering the existence of these nearly-gapless modes. We note that this phenomenon has a counterpart in non-gravitational systems and in particular shares a certain analogy with the behavior of attractive cold atoms at a quantum phase transition \cite{Dvali:2012:PhaseTransition}. 
 
  The above fully-quantum view on classical gravity is not limited to black holes. In particular, applications to cosmological solutions such as de Sitter and anti-de Sitter spaces were suggested already in 
  \cite{Dvali:2011NPortrait} and were further discussed in \cite{Dvali:2013Compositness, Dvali:2014Quantum, Kuhnel:CorpusculardS, Berezhiani:2016CorpuscularInflation}. Already simple qualitative arguments imply that -- if a consistent coherent state resolution of de Sitter metric exists -- it should come in form of a quantum state of constituent gravitons of occupation number  ${N=\hbar^{-2} M_p^2/\Lambda}$, where $\Lambda^{-1}$ is the square of the curvature radius and $M_p$ is the Planck mass. 

 A fundamental new concept which emerges in the coherent state resolution of both black holes as well of de Sitter is their {\it quantum break-time}. As already explained, this notion comes from the fact that interactions among the quantum constituents lead to a decoherence of the quantum state and thereby to a departure from the classical mean field description. The subsequent breakdown of the classical approximation after a certain finite time has important implication for any system for which the classical approximation is crucial. In particular, this implies for de Sitter that its description in terms of a classical metric cannot be future-eternal. The estimated quantum break-time is $t_{\text{q}}\sim N/\sqrt{\Lambda}$ \cite{Dvali:2013Compositness, Dvali:2014Quantum}.  We note that this time-scale appeared previously in several different contexts \cite{timeD, timeMarkkanen}, but  we shall focus on the physical meaning it acquires in our framework. The purpose of the present paper is first to develop some very general concepts of quantum breakdown of classical systems and subsequently to apply them to de Sitter to derive its quantum break-time.
 
 \subsubsection*{Outline}
 
  As a first step, we introduce the notion of classical break-time, which is entirely due to classical nonlinearities, and work out how it differs from the quantum break-time, which interests us. Subsequently, we illustrate these concepts using a simple model of an anharmonic oscillator. Finally, we discuss how it is generically possible to get an estimate of the quantum-break time of a system even if one neglects classical nonlinearities. This extrapolation is crucial for our application to de Sitter.
 
  In section \ref{sec:quantumResolution}, we shall develop a coherent state description of de Sitter by implementing the earlier approach of \cite{Dvali:2013Compositness}. Using a first-order weak-field expansion, we construct an explicit model which provides a quantum description of de Sitter as expectation value of a coherent state of high graviton occupancy. This enables us to reproduce the de Sitter metric as free-field evolution for short times. We calculate the classical break-time of our toy model, after which it deviates from the full nonlinear de Sitter solution. We conclude the section by performing some immediate  consistency checks.
  
  The fundamental quanta of our toy model are defined as spin-2 excitations on a Minkowski vacuum. This is one key point of our approach. Unlike the standard treatment, for us de Sitter is not a vacuum state, but rather a particular coherent state defined in the Fock space with Minkowski vacuum.  Correspondingly, the symmetry of de Sitter metric is {\it not}  a symmetry of the vacuum of the fundamental theory, but rather emerges as an effective symmetry of the expectation value over a particular coherent state, i.e., it is a symmetry of the mean field description.  
  
  In section \ref{sec:SemiClassicalLimit}, we show how the known classical and semi-classical processes are reproduced as full quantum evolution of the coherent state. This happens in the limit $N \rightarrow \infty$, in which any back reaction to the classical mean field can be neglected.  In this limit, the evolution of external probe particles fully reproduces the semi-classical evolution in the background classical metric. We illustrate our results with the help of two concrete phenomena, namely the redshift, which an external particle experiences in a de Sitter background, and the dilution of a gas of massive particles.
   
    But our quantum picture of de Sitter does not  only allow us to understand the quantum origin of classical and semi-classical processes. Its crucial novelty is that it captures effects which are invisible in the semi-classical approximation and which lead to a violation of de Sitter invariance and to a finite quantum break-time. We discuss those in section \ref{sec:ParticleProduction} using the example of Gibbons-Hawking particle production. In our quantum description, it arises as scattering process of the constituent gravitons.  After showing how well our simple toy model  captures the key features of standard semi-classical results, including the Boltzmann-type suppression of the production of heavy states, we estimate the quantum effects,  which lead to a departure from the classical metric description. They are of order $1/N$ as compared to the leading order mean field evolution, but are cumulative in nature and start to dominate over a sufficiently long time-scale. 
    
    Using the same extrapolation technique as for the anharmonic oscillator, we give an estimate of the quantum break-time of full nonlinear de Sitter. We show that it satisfies exactly the same universal relations as the quantum-break time in non-gravitational systems such as the model of the anharmonic oscillator. Furthermore, we demonstrate that the quantum break-time of de Sitter leads to a bound on the number of light particle species in our Universe.
 
  Finally, we will discuss in section \ref{sec:CosmologicalConstant} that our framework turns the cosmological constant problem from an issue of naturalness into a question of quantum-mechanical consistency \cite{Dvali:2014Quantum}. We explicitly estimate the maximal duration of the entire classical history of our Universe as well as the maximal value of the cosmological constant consistent with the age of our Universe.
  
  We conclude by summarizing our findings in section \ref{sec:Conclusion}.

\subsection{Classical versus quantum break-time}

There is an immediate difficulty which one encounters when attempting the quantum-corpuscular resolution of classical solutions for which nonlinear interactions are fully important, e.g., solitons and other nonlinear field configurations. This is the problem of identifying the Fock space of creation and annihilation operators $\hat{a}^{\dagger}$, $\hat{a}$ in which the coherent quantum state describing the given classical solution can be defined. 
 
If for a given solution the nonlinear interactions are important, in general its quantum constituents $\hat{a}^{\dagger}$, $\hat{a}$ differ from the quantum constituents $\hat{a}_{\text{free}}^{\dagger}$, $\hat{a}_{\text{free}}$ which describe the (almost) free waves obtained by solving the classical equations of motions in the same system, but in a weak field limit.  The obvious reason for this difference is that in nonlinear waves, interactions are important and the would-be free particles are off-shell. Correspondingly, the dispersion relation of quanta $\hat{a}^{\dagger}$, $\hat{a}$ is in general very different from $\hat{a}_{\text{free}}^{\dagger}$, $\hat{a}_{\text{free}}$. In other words, $\hat{a}_{\text{free}}^{\dagger}$ creates a free quantum whereas $\hat{a}^{\dagger}$ creates one which interacts with the other background constituents.
 
The idea is to choose the operators $\hat{a}^{\dagger}$, $\hat{a}$, which take into account all nonlinear interactions, in such a way that a coherent state formed out of them leads to the correct classical expectation value for all times, as long as one neglects how the quantum state evolves. For a generic classical solution, however, it is impossible to find such operators. To overcome this problem, our strategy is to approximate the classical solution so that appropriate operators can be found for this modified classical function. It is crucial to note that this approximation is purely classical: We replace the exact classical solution by a different classical function. For the approximate function, we find an exact quantum-corpuscular resolution in terms of $\hat{a}^{\dagger}$, $\hat{a}$. When the description of the classical solution in terms of $\hat{a}^{\dagger}$, $ \hat{a}$ breaks down, this only happens because the approximation of the exact solution on the classical level stops being valid. Consequently, we define the time scale after which the exact and approximate classical solution deviate significantly as {\it classical break-time} $t_{\text{cl}}$. Put simply, the classical break-time $t_{\text{cl}}$ is the time-scale 
after which classical nonlinearities significantly correct the classical free field approximation.  

In the quantum theory, however, a second time-scale arises. As already explained, this happens because nonlinearities (or interactions with another field) enable the constituents of the coherent state to scatter and thus trigger a quantum evolution of the state. Each rescattering slightly spoils the coherence of the state in such a way that the expectation value taken over the time-evolved quantum state no longer matches the time evolution of the classical field. A significant departure from the classical solution takes places as soon as a significant fraction of particles has rescattered. We shall call this time scale the {\it quantum break-time} $t_{\text{q}}$. It is important to note the quantum break-time is inaccessible in the classical limit, which corresponds to an infinite occupation number in the coherent state.
  
It is crucial not to confuse the two time-scales introduced above. The quantum break-time $t_{\text{q}}$ describes the point when quantum evolution departs from the classical solution, i.e., the expectation values over the coherent state no longer evolve according to classical equations. In particular, this happens when the state can no longer be described as a coherent state. In contrast, the classical break-time $t_{\text{cl}}$ corresponds to the time-scale after which nonlinearities correct the evolution of the system, while it remains perfectly classical. The distinction and separation of these two concepts will play a key role in interpreting our findings.
  
We will specialize to systems which can be described in terms of only two relevant quantum numbers:\\
   1)  the number $N$ which measures the occupation number of quanta in the corresponding coherent state; and\\   
    2) the typical dimensionless quantum coupling constant 
   $\alpha$ which measures the strength of interaction among the quanta, i.e., the strength of the off-diagonal and higher order terms in the Hamiltonian. \\   
 Since we  are dealing with field configurations which can be described classically in a first order approximation, we shall always work with large $N$. Moreover, we have $\alpha \ll 1$ in all the systems of our interest. In such a case, the useful parameter turns out to be the collective coupling $\alpha N$. When $\alpha N \ll 1$, then both classical nonlinear as well as the quantum effects can be ignored for certain times. This means that we can approximate the exact classical solution by a free, non-interacting solution. Consequently, we approximate the quantum constituents as free: $\hat{a}^{\dagger} = \hat{a}_{\text{free}}^{\dagger}$. For $t<t_{\text{cl}}$ and $t<t_{\text{q}}$, this implies that we can view the classical solution as coherent state of free quanta.

  \subsection{An explicit example}
  
  \subsubsection*{Classical break-time}
  In order to clearly identify the effect we are after, let us explicitly compare the classical and quantum break-time in a concrete example. Consider a scalar field $\phi$ in $3+1$ dimensions with the Lagrangian, 
\begin{equation}   
   \mathcal{L} = {1 \over 2}  \partial_\mu \phi  \partial^\mu \phi  - 
   {1 \over 2} \omega^2 \phi^2  -  {1 \over 4}  \lambda^2 \phi^4 \,, 
   \label{class}
   \end{equation} 
  where $\omega^2$ and $\lambda^2$ are positive parameters. 
  Since we need to confront the classical and quantum time-scales, it is useful to keep for the time being $\hbar$ explicit while setting the speed of light equal to one
  throughout the paper.  Then, in the classical theory, the dimensions of the parameters are 
  $[\omega^{-1}] = (time)$ and $[\lambda^{-2}] =  (time) \times (energy)$, whereas the dimensionality of the field 
  is $[\phi] = \sqrt{(energy)/(time)}$.  
   
 The classical equation of motion, 
  \begin{equation}   
   \Box \phi   +   \omega^2 \phi    +   \lambda^2 \phi^3  = 0\,,  
   \label{classEQ}
   \end{equation}     
  admits a purely time-dependent solution $\phi(t)$ which describes anharmonic oscillations around the value $\phi = 0$ with an amplitude $A$. For small nonlinear coupling $\lambda^2$ and a small amplitude $A \ll \omega/\lambda$, we can ignore the nonlinear term and approximate the solution by the one of a free theory, i.e., by the harmonic oscillations $\phi_{\text{(l)}}(t) \, = \, A \cos(\omega t)$. This approximation, however, is only valid on the time-scale of the classical break-time:
  \begin{equation}    
  t_{\text{cl}} \, \equiv {\omega \over \lambda^2 A^2}  \,.
  \label{tclass} 
  \end{equation}
  Afterwards, the classical effect of a nonlinear interaction becomes large and must be taken into account. As expected, $t_{\text{cl}}$ contains no dependence on $\hbar$ as it only measures the strength of classical nonlinear effects.    
 
 \subsubsection*{Quantum break-time due to rescattering}
 What interests us, however, it the quantum break-time. In order to determine it, we need to understand the time-dependent oscillating scalar field in a quantum language. As explained above, the approximation as free solution enables us to do so since we can use free quanta as constituents of our quantum state. As the expectation value is furthermore translation-invariant, we only need to consider ones with zero momentum.
In the language of a quantum theory, the 
 oscillating classical field $\phi_{\text{(l)}}(t)$ consequently corresponds to the expectation value of a quantum field $\hat{\phi}_{\text{(l)}}$, 
 \begin{equation} 
   \phi_{\text{(l)}}(t) = \langle N | \hat{\phi}_{\text{(l)}} | N \rangle  \,,  
   \label{expN}
\end{equation}
  over a coherent state $| N \rangle$ of mean occupation number-density $n = {1 \over \hbar} \omega A^2$, or equivalently, mean occupation number $N =  {1 \over \hbar} {A^2 \over \omega^2}$ (per volume $\omega^{-3}$).\footnote
 {
     The classical expectation value, $\phi_{\text{(l)}}(t) \, = \, A \cos(\omega t)$, determines the number density:
     \begin{align*}
		  \langle N | \hat{\phi}_{\text{(l)}} | N \rangle =  \langle N | \int \! \diff^3 \vec{k} \, \frac{1}{2\sqrt{\omega}} \left(\hat{a}^{\dagger}\, \text{e}^{i\hbar^{-1}kx} + \hat{a}\, \text{e}^{-i\hbar^{-1}kx}\right) | N \rangle =   \sqrt{\frac{\hbar N}{\omega V}} \cos(\omega t) \,,
     \end{align*}	
  where $V$ is the volume.
  } 

In this model, nonlinearities do not only lead to a classical break-time, i.e., a departure of  $\phi(t)$ from harmonic oscillations. More importantly, they cause a finite quantum break-time since the quantum scattering leads to a decoherence of the coherent state $| N \rangle$.
 The rate of this quantum scattering is controlled by the quantum coupling constant $\alpha$, which in the quantized theory of (\ref{class}) is defined as $\alpha \equiv \hbar \lambda^2$.  In particular, the  dimensionless scattering amplitude corresponding to a $2 \rightarrow 2$  process is given by $\alpha$.

 Before estimating the quantum break-time, it is instructive 
 to rewrite the classical break-time in terms of quantum 
 parameters $\alpha$ and $N$. We get 
 \begin{equation}    
  t_{\text{cl}} \, \equiv {1 \over \omega (\alpha N)}  \,.
  \label{tclassQ} 
  \end{equation}
   The correct classical limit corresponds to taking $\hbar \rightarrow 0$, while keeping all the classical parameters  (such as $\omega, \lambda$ and $A$) finite. It is obvious that in this limit,  we have a double-scaling behavior such that $\alpha \rightarrow 0,~N \rightarrow \infty$, but the collective coupling $\alpha N$ stays finite.  Thus, it is clear from expression (\ref{tclassQ}) that the classical break-time is controlled by the collective coupling $\alpha N$. 
 
 We shall now turn to estimating the quantum break-time in this system. The rate at which a fixed pair of particles scatters per volume $\omega^{-3}$ is given by:
  \begin{equation}  
    \Gamma \sim \omega \alpha^2  \,. 
    \label{RATE}
    \end{equation}
 Taking into account that the approximate number of pairs within the volume of interest is $N^2$, we obtain the quantum break-time as the minimal time needed till a fraction of order one of the particles has experienced scattering:
 \begin{equation}
   t_{\text{q}} \sim N\, (N^2 \Gamma)^{-1} \sim   {1 \over \omega \alpha^2 N} \,. 
   \label{Qbreak}
   \end{equation}  
  
  We note that we can express the relationship between the quantum and classical break-times in the following form, 
  \begin{equation}
  t_{\text{q}} = {t_{\text{cl}} \over \alpha} \, ,  
  \label{2times}
  \end{equation} 
 which shows that at weak quantum coupling, the classical effects of nonlinearities become important faster than their quantum effects.  For $\hbar \rightarrow 0$, the quantum break-time becomes infinite, as it should. 
 
 \subsubsection*{Quantum break-time and particle decay}
 
  We would like to investigate the influence of particle decay on the quantum break-time. For this, we shall enlarge the theory (\ref{class}) by introducing a coupling of $\phi$ to a new particle species $\psi$ to which $\phi$ can decay. The Lagrangian becomes,    
  \begin{equation}   
   \mathcal{L} = {1 \over 2}  \partial_\mu \phi  \partial^\mu \phi  - 
   {1 \over 2} \omega^2 \phi^2  -  {1 \over 4}  \lambda^2 \phi^4 \, 
  + \, i \bar{\psi} \gamma_{\mu} \partial^{\mu} \psi   -  \lambda \phi \bar{\psi}\psi  \, .
   \label{classNEW}
   \end{equation} 
   We took $\psi$ to be a massless fermion, although the spin and the mass are unimportant for this consideration as long as the new particle is light enough to allow for a decay of  $\phi$. In order to make a clear comparison, we have taken the coupling to fermions of the same strength as the bosonic self-coupling. In this way, the amplitudes of both $2\rightarrow 2$ scatterings,  $\phi + \phi \rightarrow  \phi + \phi$ and $\phi + \phi \rightarrow  \psi + \psi$,  are controlled by the same amplitude $\alpha$.  

 The decay rate of a $\phi$-particle into a pair of $\psi$-quanta is given by $\Gamma_{\text{decay}} \sim \omega \alpha$. The corresponding half-life time, $t_{\text{decay}} = \Gamma_{\text{decay}}^{-1}$, is related to the previously-derived time-scales as 
\begin{equation} 
 t_{\text{decay}} \sim t_{\text{cl}} N \sim  t_{\text{q}} (\alpha N)\, .
 \label{3times} 
 \end{equation} 
 Since we have $N \gg 1$ in the classical regime, $t_{\text{decay}}$ is always longer than $t_{\text{cl}}$, as it should. In turn, the above expression creates the impression that for $\alpha N \ll 1$, the quantum decay time $t_{\text{decay}}$ can be much shorter than the quantum break-time (\ref{Qbreak}) obtained by rescattering of $\phi$-quanta. 
 
This turns out not to be completely correct. Of course, the quantum 
two-body decay affects the  evolution of the oscillating scalar field $\phi(t)$ on the time-scale $t_{\text{decay}}$. It does not, however, affect the classicality of the solution. If we only take into account the 
decays of individual quanta, ignoring interaction effects such as rescatterings or recoils, the evolution is still well-described by a classical field which satisfies a  modified classical equation with a friction term, 
 \begin{equation}   
   \ddot{\phi}   +  \Gamma_{\text{decay}}\dot{\phi} +  \omega^2 \phi    +   \lambda^2 \phi^3  = 0\,.   
   \label{classEF}
   \end{equation}     
 The reason for the validity of classical evolution is that the decay process alone (without taking into the account back reaction)  does not affect the coherence of the state of $\phi$-quanta. Indeed, the part of the interaction Hamiltonian which is responsible for a two-body decay of $\phi$-quanta only contains the annihilation operator $\hat{a}$ of $\phi$-quanta and the creation operators for $\psi$-particles. The coherent state $|N\rangle$, however, is an eigenstate of $\hat{a}$ and hence unchanged under its action. Therefore, a pure decay process does not affect the quantum break-time $t_{\text{q}}$. In order to violate the coherence of the state, the interaction term must also contain creation operators of $\phi$-particles, i.e., it must describe a re-scattering process. Since the process $\psi \rightarrow \psi + \phi$ is kinematically forbidden, the leading order process is $\psi \rightarrow \phi + \phi + \psi$, whose amplitude scales like $\alpha$. Thus, the rate goes like $\alpha^2$ and decay processes do not lead to a shorter quantum break-time. Even in the presence of particle decay, the quantum break-time is given by the expression (\ref{Qbreak}). 
  
 \subsection{Extrapolation of the quantum break-time}
 
  As we have seen, the quantum break-time is generically bigger than the classical one. Since the linear approximation, on which we base our toy model, already breaks down on the time-scale of the classical break-time, one is immediately led to wonder whether it is possible to make any statement at all about the quantum break-time of the full nonlinear system.  We can split the essence of this question in two parts:\\
  I) Does the toy model capture the difference between quantum and classical effects? \\
  II) How toy is the toy model, i.e., what properties does it share with the full nonlinear theory?
  
  Clearly, the answer to part I) is positive. The linear toy model explicitly distinguishes between classical and quantum effects since they depend on independent and fundamentally different parameters. The classical break-time is controlled by the collective coupling $\alpha N$ (see equation \eqref{tclassQ}), which is a classical quantity independent of $\hbar$. In contrast, the quantum effects are determined by $\alpha$ (see equation \eqref{Qbreak}), which vanishes in the classical limit $\hbar \rightarrow 0$. Therefore, we can vary the respective strength of the effects independently. This reflects the fact that the two effects -- classical nonlinearities and quantum scatterings -- are of fundamentally different nature.
  
  This observation is the key to answering part II): In order to promote our linear toy model to the full solution, we would have to take into account the effect of classical nonlinearities. This would amount to changing the Fock basis by performing an appropriate Bogoliubov transformation and to using the coherent state description in terms of interacting quanta $\hat{a}^{\dagger}$, $\hat{a}$ instead of free ones $\hat{a}_{\text{free}}^{\dagger}$, $\hat{a}_{\text{free}}$. In this way, we could increase the classical break-time. But even in a general nonlinear description, quantum scattering events which lead to decoherence of the initially coherent state continue to occur. They still lead to a gradual departure from the classical evolution on a time-scale suppressed by the coupling $\alpha$.\footnote{
  	Putting it in the language of large-$N$ physics, the quantum effects, which are controlled by $\alpha$, can be studied as series in $\alpha$-expansion for any given order in $\alpha N$. 
  }
  Therefore, it is reasonable to extrapolate the quantum break-time, which we calculated in the toy model, to get an estimate of the quantum break-time in the full nonlinear theory. This procedure yields the correct order of magnitude as long as some coherent state description of the full nonlinear theory exists.
 
Having gained intuition in this model of anharmonic oscillations, we will now turn to de Sitter and the study of its quantum break-time. As we shall see below, the quantum resolution of de Sitter space amounts to $\alpha N = 1$.  Hence, for this case, the quantum effects are exactly $1/N$-corrections to classical nonlinear effects. In full analogy to the treatment of the anharmonic oscillator, we will calculate the break-time for a linear toy model and then extrapolate it to get a corresponding estimate for the full nonlinear theory. For this to work, the only physical assumption which we need is that some quantum description of de Sitter exists.

\section{A quantum description of the de Sitter metric}
\label{sec:quantumResolution}

\subsection{Short-time descriptions for classical de Sitter}

\subsubsection*{Linear gravity with constant source}
Our first goal is to find a quantum description of de Sitter space. As we have seen in the case of the anharmonic oscillator, this means that we have to find a classical approximation to the de Sitter solution such that an exact quantum resolution can be found for this approximate classical solution. To this end, we shall follow \cite{Dvali:2007Degravitation}, where a short-time description of the de Sitter metric was obtained as exact solution in a theory of linearized Einstein gravity with a cosmological constant source $\Lambda$ as source.\footnote
{
   Note that we corrected minor numerical factors as compared to \cite{Dvali:2007Degravitation}.
}
 
First we linearize the Einstein equations {\it on top of the Minkowski metric} $\eta_{\mu\nu}$:
\begin{align}
	\epsilon_{\mu\nu}^{\alpha\beta}\widetilde{h}_{\alpha\beta}  = -2\Lambda \eta_{\mu\nu}\,, \label{eqn:Cla:MasslessGravitonEomUngauged}
\end{align}
with the linearized Einstein tensor defined as  $\epsilon_{\mu\nu}^{\alpha\beta}h_{\alpha\beta} \equiv
	\Box h_{\mu\nu} - \eta_{\mu\nu}\Box h - \partial_\mu\partial^\alpha h_{\alpha\nu} -\partial_\nu\partial^\alpha h_{\alpha\mu} + \partial_\mu \partial_\nu h + \eta_{\mu\nu}\partial^\alpha\partial^\beta h_{\alpha\beta}$.  The gauge symmetry is, 
	\begin{equation} 
	\widetilde{h}_{\mu\nu} \rightarrow  \widetilde{h}_{\mu\nu}  \,  + \, 
	\partial_{\mu}\xi_{\nu}  +  \partial_{\nu}\xi_{\mu} \,, 
	\label{lineargauge}
	\end{equation} 
	where $\xi_{\nu}$ is a gauge-transformation parameter. In de Donder gauge, $\partial^\mu \widetilde{h}_{\mu\nu} = \frac{1}{2} \partial_\nu \widetilde{h} $, the equation takes the following form:
\begin{align}
\Box \left(\widetilde{h}_{\mu\nu} - \frac{1}{2} \eta_{\mu\nu} \widetilde{h}\right) = -2\Lambda \eta_{\mu\nu} \,. \label{eqn:Cla:MasslessGravitonEom}
\end{align}
 Here $\widetilde{h}_{\mu\nu}$  denotes  a small departure from the Minkowski metric caused by the presence of a constant source $\Lambda$. Correspondingly, $\widetilde{h}_{\mu\nu}$ is dimensionless,  whereas $\Lambda$ has a dimensionality of frequency-squared. In the full nonlinear de Sitter solution, $\Lambda$ would correspond to the de Sitter Hubble parameter, $\Lambda \, =  \, 3 H^2$, which is equivalent to the curvature radius $R_H = H^{-1}$.
 
It is very important not to confuse $\widetilde{h}_{\mu\nu}$ with the  linear metric perturbation on top of a de Sitter metric: $\widetilde{h}_{\mu\nu}$ is a short-time approximation of the de Sitter metric itself, not a fluctuation on top of it!

Following the results of \cite{Dvali:2007Degravitation}, the equations of motion \eqref{eqn:Cla:MasslessGravitonEom} are solved by
\begin{subequations} \label{eqn:Cla:MasslessSolution}
\begin{align}
	\widetilde{h}_{00} =& \Lambda t^2 \,,\\
	\widetilde{h}_{0i} =& -\frac{2}{3}\Lambda t x_i \,,\\
	\widetilde{h}_{ij} =& -\Lambda t^2 \delta_{ij} - \frac{\Lambda}{3}\epsilon_{ij}\,,  
\end{align}
\end{subequations}
where $\epsilon_{ij} = x_ix_j$ for $i\neq j$ and $0$ otherwise. Still following \cite{Dvali:2007Degravitation}, we apply a diffeomorphism to obtain
\begin{align}
\diff s^2 = \diff t^2 - \left(1+\frac{1}{3}\Lambda t^2   \right)\delta_{ij} \diff x^i \diff x^j - \frac{1}{3} \Lambda x_i x_j \diff x^i \diff x^j \,,
\label{LDSFRW}
\end{align}
which is an approximation of a de Sitter Universe in 
closed Friedmann-Robertson-Walker slicing:
\begin{align}
\diff s^2 = \diff t^2 -\cosh^2\left(\sqrt{\Lambda/3}\, t\right)\left(\frac{\diff r^2}{1-\Lambda r^2/3} + r^2 \diff \Omega^2 \right)\,.
\label{DSFRW}
\end{align}
 To first order in the $\Lambda t^2$- and 
 $\Lambda r^2$-expansion, it is clear that \eqref{DSFRW} reproduces 
\eqref{LDSFRW}. We conclude that our approximation of weak gravity is valid as long as $t \ll \Lambda^{-\frac{1}{2}}$ and $x \ll \Lambda^{-\frac{1}{2}}$. This yields the classical break-time 
\begin{align}
t^{(0)}_{\text{cl}} =  \Lambda^{-\frac{1}{2}} \, .
\label{dbreak} 
\end{align} 

It is instructive to confront expression \eqref{dbreak} for the classical break-time of de Sitter with its counterpart for anharmonic oscillations of the scalar field  \eqref{tclass}.  The classical oscillations of the scalar field are defined by three parameters: frequency $\omega$, amplitude $A$ and the strength of classical nonlinear interaction 
$\lambda^2$. Let us define corresponding parameters for the classical de Sitter space at short times. The characteristic time dependence is defined by the Hubble scale, $\omega_{\text{grav}} = \sqrt{\Lambda}$. Secondly, the amplitude  of the canonically  normalized graviton field beyond which nonlinearities become crucial is given by the inverse square root of Newton's constant, $A_{\text{grav}} = 1 / \sqrt{G_N}$. Finally, the strength of gravitational nonlinearities in de Sitter is defined by 
$\lambda^2_{\text{grav}} =  \Lambda G_N$. Replacing in the expression for the classical  break-time of the anharmonic scalar field \eqref{tclass} 
the parameters $\omega, A$ and $\lambda^2$ by their gravitational counterparts $\omega_{\text{grav}}, A_{\text{grav}}$ and $\lambda_{\text{grav}}^2$, we get exactly expression \eqref{dbreak}, which we had obtained by comparing the metrics \eqref{LDSFRW} and \eqref{DSFRW}. Despite the fact that the anharmonic scalar field and de Sitter are very different systems, their classical break-times obey the universal relation \eqref{tclass}. The difference is that in gravity, unlike in the scalar field case, only two out of the three parameters are independent. Correspondingly, we do not have the flexibility of making the classical break-time longer than $1/\sqrt{\Lambda}$.   

\subsubsection*{Mapping de Sitter on a graviton mass}
Our subsequent task is to provide a quantum-corpuscular description of 
\eqref{eqn:Cla:MasslessSolution} in form of a coherent state. To this end, we slightly deform the linearized theory by promoting it into a relative theory, which has exactly the same metric description for the relevant  time-scales, but for which the coherent state description is much more straightforward. The choice of deformation for the linear theory is unique.  It is given by the only existing ghost-free linear theory of a spin-2 field beyond linearized Einstein gravity: Pauli-Fierz theory of massive spin-2.
 	
But what are the reasons for adding a mass? First, as observed in \cite{Dvali:2007Degravitation}, this deformation leads to a solution which reproduces the de Sitter metric for times $t \ll m^{-1}$ even in the absence of the cosmological constant term.  Here $m$ is the mass of the gravitational field, but since we are still in a classical theory, it has a dimensionality of frequency. For short time-scales, the graviton mass consequently has the same effect as the cosmological term. This means that observers coupled to such a gravitational field, for a short time-scale cannot tell whether they live in a de Sitter metric of Einstein theory or in a coherently oscillating field of a massive Fierz-Pauli graviton on top of a flat Minkowski vacuum. Since we want to map the cosmological constant to a graviton mass, we already expect at this point that
\begin{align}
m \approx \sqrt{\Lambda}\,. 
\end{align}	
We will elaborate on this relation later but keep $m$ arbitrary for now. 

Secondly, such a deformation allows for a simple coherent state interpretation of the de Sitter metric: It is much more straightforward to describe a coherently-oscillating free massive spin-2 field as the quantum coherent state than its massless counterpart sourced by a cosmological term.   
 
  In addition, it matches the physical intuition that if de Sitter space allows for a sensible corpuscular resolution in form of a coherent state, the constituents must have frequencies given by the Hubble parameter  since this is the only scale of the classical geometry. Thus, these constituents can be viewed as some sort of {\it off-shell} gravitons of non-zero frequencies set by $H$. The mass term is the simplest terms which provides such an effective off-shell dispersion relation. Thus, for a sufficiently short time-interval, $t \ll t_{\text{cl}}$, we can think of the gravitons of the massless theory which are put off-shell by nonlinearities as on-shell massive gravitons of a free theory. This mapping allows for the coherent state interpretation of  de Sitter metric for sufficiently small times. Although the approximation breaks down after $t_{\text{cl}}$, it suffices to "fish out" the $1/N$-quantum effects which lead to a departure from the coherent state picture.   

 We therefore modify our theory by adding a graviton mass $m$ and removing the cosmological constant source. To linear order, the massive graviton, which we shall denote as $h_{\mu\nu}$, obeys:
  \begin{align}
 \epsilon_{\mu\nu}^{\alpha\beta}h_{\alpha\beta} + m^2(h_{\mu\nu}-\eta_{\mu\nu}h) =  0\,, \label{eqn:Cla:MassiveGravitonEom}
 \end{align}
with the linearized Einstein tensor given as above. Additionally, it must satisfy the Fierz-Pauli constraint
\begin{align}
 	\partial^\mu(h_{\mu\nu} - \eta_{\mu\nu} h) \, = \, 0\,,   
 	\label{PFconstr} 
\end{align} 
 which shows that it propagates five degrees of freedom. Following \cite{Dvali:2007Degravitation, Dvali:2006ModifiedGravity}, these degrees of freedom  can be split according to irreducible massless representations of the Poincar\'{e} group into three different helicity components: helicity-2 $\widetilde{h}_{\mu\nu}$, helicity-1 $A_\mu$ and helicity-0 $\chi$:  
 	\begin{align}
 	h_{\mu\nu} = \widetilde{h}_{\mu\nu} + {1 \over m} (\partial_\mu A_\nu + 
 	\partial_\nu A_\mu ) + 	
 	\frac{1}{6} \eta_{\mu\nu}\chi - \frac{1}{3} \frac{\partial_\mu \partial_\nu}{m^2} \chi \,. \label{eqn:Cla:MassiveGravitonSplit}
 	\end{align}
 	This decomposition is unique in the sense that in this basis, the kinetic mixing among different helicities is absent and in the limit $m\rightarrow 0$, the field $h_{\mu\nu}$ "disintegrates" into three independent massless representations of the Poincar\'{e} group: spin-2, spin-1 and spin-0.  Notice that the gauge redundancy (\ref{lineargauge}) of the massless  theory is not lost.  The gauge shift  (\ref{lineargauge})  is compensated by a corresponding shift of $A_{\mu}$,
 \begin{equation}
 	A_{\mu} \rightarrow  A_{\mu}\,  - \, m\xi_{\mu}\,.    
 	\label{shift} 
 \end{equation} 
 	Hence, $A_{\mu} $ acts as St\"uckelberg-field, i.e., we can continue enjoying the gauge freedom for fixing the gauge of the  
 	$\widetilde{h}_{\mu\nu}$-component. This is particularly useful, since  following \cite{Dvali:2006ModifiedGravity}, we can integrate out the additional helicities and write down an effective equation for $\widetilde{h}_{\mu\nu}$.  
 
 This equation in de Donder gauge, $\partial^\mu\widetilde{h}_{\mu\nu} = \frac{1}{2} \partial_\nu \widetilde{h}$, is a massive wave equation:
 \begin{align}
 (\Box +m^2)\left(\widetilde{h}_{\mu\nu} - \frac{1}{2} \eta_{\mu\nu} \widetilde{h}\right) = 0\,.  \label{eqn:Cla:MassiveHelicity2Eom}
 \end{align}
 One solution is given by
 \begin{subequations} \label{eqn:Cla:MassiveHelicity2Solution}
 	\begin{align}
 	\widetilde{h}_{00} =& - \frac{2\Lambda}{m^2}\cos(mt)\,,\\
 	\widetilde{h}_{0i} =& \frac{-2\Lambda}{3m}\sin(mt)x_i\,,\\
 	\widetilde{h}_{ij} =& \frac{2\Lambda}{m^2}\cos(mt)\delta_{ij} - \frac{\Lambda}{3}\cos(mt)\epsilon_{ij}\,,
 	\end{align}
 \end{subequations}
 where additional helicity-0 part assumes the following form: 
 \begin{align}
 \chi = -\widetilde{h} = \frac{8 \Lambda}{m^2} \cos(mt) \,. \label{eqn:Cla:Chi}
 \end{align}
 This formula is a manifestation of the fact that $\chi$ and $\widetilde{h}$ are not independent but mix through the mass term and undergo simultaneous coherent oscillations. Correspondingly, the oscillating classical field represents a coherent state composed out of quanta which reside both in $\chi$ and $\widetilde{h}$.
 
For $t \ll m^{-1}$, the oscillating solution 
 (\ref{eqn:Cla:MassiveHelicity2Solution})
 of the massive theory without any cosmological constant fully reproduces  -- up to an additive constant --  the de Sitter solution \eqref{eqn:Cla:MasslessSolution} of the massless theory with a cosmological constant as source:
 \begin{equation} 
 \widetilde{h}_{\mu\nu}^{(m\neq 0,\ \Lambda=0 )} =  
 \widetilde{h}_{\mu\nu}^{(m=0,\ \Lambda\neq 0)} - 
 \eta_{\mu\nu} \frac{2\Lambda}{m^2} \, .
 \label{Mrelation} 
 \end{equation}
 Due to the normalization of the amplitude $\frac{2\Lambda}{m^2}$, this relation holds irrespective of the graviton mass. Obviously, the classical break-time after which the oscillating field no longer approximates the massless solution (\ref{eqn:Cla:MasslessSolution}) is 
 \begin{equation}
 t_{\text{cl}} =  m^{-1} \, .
 \label{dbreakM} 
 \end{equation} 
 
Finally, we choose $m$ such that the classical break-times of linearized solutions in the two theories are the same: in nonlinear massless gravity, the classical break-time is reached  when the variation of the dimensionless metric $\widetilde{h}_{\mu\nu}$ becomes of order one. This is equivalent to the statement that the linearized approximation breaks down when the canonically-normalized field $\widetilde{h}_{\mu\nu}$ becomes of order of the Planck mass $M_P$.  Applying the same criterion to the oscillating solution of the linearized massive theory, we must set 
 \begin{align}
 m \approx \sqrt{\Lambda}\,. \label{eqn:Cla:ApproximateMass}
 \end{align}	
 With this choice, the classical break-times in the two theories match: the time-scale (\ref{dbreak}) of the  breakdown -- due to classical nonlinearities -- of the linearized de Sitter solution  (\ref{eqn:Cla:MasslessSolution}) of a massless Einstein theory and the time-scale (\ref{dbreakM}) of the breakdown of the same solution (\ref{eqn:Cla:MassiveHelicity2Solution}) -- due to the mass term -- in the linear  Pauli-Fierz massive theory.  This means that the classical nonlinearities in Einstein gravity  and the mass term in the free Pauli-Fierz theory are doing the same job of putting the solution (\ref{eqn:Cla:MasslessSolution}) out of business.  In section \ref{sec:ResolutionDiscussion}, we will see another reason why we have to fix the mass according to \eqref{eqn:Cla:ApproximateMass}. 
 
We conclude that we have accomplished our first goal of finding an appropriate approximation to the exact classical solution. Its classical break-time is
 \begin{align}
 t_{\text{cl}} = \Lambda^{-\frac{1}{2}} \,. \label{eqn:Cla:classicalBreakTime}
 \end{align}
 Thus, for short time-scales $t< t_{\text{cl}}$, the graviton mass fully replaces the effect of the cosmological term. This fact shall allow us to give a well-defined coherent state representation of de Sitter space during that time span.
 
 We would like to stress that although we borrow the setup of 
 \cite{Dvali:2007Degravitation}, we do not use its approach of killing 
  (i.e., de-gravitating) the cosmological constant by means of the graviton mass.
  Instead, we manufacture the de Sitter metric by replacing the cosmological constant by a graviton mass. 
  
  In this respect, the idea of our simple model is also reminiscent of the idea of 
  {\it self-acceleration} \cite{deffayet} in which a de Sitter-like 
  solution is achieved without the cosmological constant source, due to modification 
  of the graviton dispersion relation. More recently, such solutions were obtained \cite{massiveCosmology} in the theory of massive gravity of \cite{deRham}.  In the present paper, since our focus is 
  not a modification of Einstein theory but rather the creation of a simple setup which allows for coherent state interpretation of the de Sitter metric, the graviton mass is merely a computational device which replaces the effect of nonlinearities.  In other words, we map the interacting off-shell massless gravitons onto free massive ones.  Hence, we shall not be concerned with nonlinear completions of the massive theory.\footnote
  {
  	It could nevertheless be an interesting independent project to extend our quantum analysis in a full nonlinear massive theory of the type \cite{deRham}.
  }

 \subsection{The basic model} 
 
Let us summarize the basic model we shall be working with throughout the paper: We replace the linearized massless theory of an Einstein graviton $\widetilde{h}^{\mu\nu}$ coupled to a cosmological constant source
$\Lambda$,  
\begin{align}
{\mathcal L_{\text{E}}} \, = \frac{1}{16\pi}\left( \, {1 \over 2} \widetilde{h}^{\mu\nu} \epsilon_{\mu\nu}^{\alpha\beta}\widetilde{h}_{\alpha\beta} + \frac{2}{ \sqrt{G_N}}\, \widetilde{h} \, \Lambda \, +\,
 16 \pi \sqrt{G_N}\, \widetilde{h}_{\mu\nu} T^{\mu\nu} (\Psi) \, + \ldots \right) \,, 
\label{masslessTH}
\end{align}
by a linear theory of a Fierz-Pauli graviton of mass $m = \sqrt{\Lambda}$: 
\vspace{2.4\baselineskip}
\begin{align}
{\mathcal L_{\text{FP}}}  =  \frac{1}{16\pi}\left({1 \over 2} h^{\mu\nu}	\epsilon_{\mu\nu}^{\alpha\beta}h_{\alpha\beta}  +  { 1 \over 2} \Lambda (h_{\mu\nu} h^{\mu\nu} - h^2)  +    
16\pi \sqrt{G_N}\,\tilde{h}_{\mu\nu} T^{\mu\nu} (\Psi)  +  \ldots \right) \,.  
\label{massiveTH}
\end{align} 
\par \vspace{1.7\baselineskip}
This is the theory on which we will base our analysis. It is crucial to note that it does not include a cosmological constant-term. Instead, it only contains the Fierz-Pauli mass term $(h_{\mu\nu} h^{\mu\nu} - h^2)$.
Notice that for the convenience of our analysis, we only couple the Einsteinian spin-2 helicity $\tilde{h}_{\mu\nu}$ to the external source.  

 Again, by no means should one think that the cosmological term gives the graviton a fundamental mass. This is not the case as it is obvious already from counting the number of degrees of freedom. We use the fact that the Einsteinian spin-2 helicity component $\tilde{h}_{\mu\nu}$ of the Fierz-Pauli  massive graviton {\it without} cosmological term has the same form 
  as it would have in a massless theory with cosmological constant. 
 
In both Lagrangians, we moved to canonically normalized classical fields by dividing by $\sqrt{G_N}$.
Therefore, both $h_{\mu\nu}$ and $\widetilde{h}_{\mu\nu}$ 
as well as the $A_{\mu}$- and $\chi$-components of $h_{\mu\nu}$ have 
dimensionality of  $\sqrt{(energy)/(time)}$. 
 Correspondingly, the helicity decomposition of $h_{\mu\nu}$ continues to have the form given by \eqref{eqn:Cla:MassiveGravitonSplit}.  $T^{\mu\nu} (\Psi)$ denotes the 
energy momentum tensor of all modes $\Psi$ which do not belong to the mode-decomposition of the background  de Sitter solution, 
i.e., the quanta which in the quantum picture are not part of the coherent state description of the de Sitter metric. We only keep the lowest order dependence of $T^{\mu\nu} (\Psi)$ on the fields. 
 
For definiteness, we assume that  $T^{\mu\nu}$ is the stress-energy tensor of some external particles and does not contain a graviton part. Including ${h}_{\mu\nu}$ into $T^{\mu\nu}$ would result in nonlinear self-interactions of ${h}_{\mu\nu}$, which would contribute to both classical and quantum break-times.  Since our approach is to replace the effect of the self-coupling by an effective graviton mass, we will not include gravitons in $T^{\mu\nu}(\Psi)$. In addition to that, it is convenient to couple gravity exclusively to external particles with initial occupation number equal to zero because such particles manifest themselves only via quantum processes and this is precisely what we are after. 
 
 Moreover, we shall couple $T^{\mu\nu} (\Psi)$ only to the Einstein spin-2 helicity component $\widetilde{h}_{\mu\nu}$. The reason why we do not couple the $\chi$-component to external sources is that we want the external particles $\Psi$  to experience -- in the classical 
limit and during the time-scale of validity of (\ref{eqn:Cla:MassiveHelicity2Solution}) -- "life" in an effective de Sitter metric.\footnote
   {
   	Notice that the coupling of the helicity-1 
 component $A_\mu$ to $T_{\mu\nu}$ automatically vanishes due to the 
 conservation of the source: $\partial^\mu T_{\mu\nu} = 0$.
 }
As explained above, the solution \eqref{eqn:Cla:MassiveHelicity2Solution} for the helicity-2 part $\widetilde{h}_{\mu\nu}$ alone suffices for that since it is equivalent to the de Sitter solution (\ref{eqn:Cla:MasslessSolution}) of a linearized massless theory for $t<t_{\text{cl}}$.
 
Finally, we further split the helicity-2 component $\widetilde{h}_{\mu\nu}$ according to the symmetries of the Poincar\'{e} group. 
It suffices for our conclusions to focus exclusively on the scalar part, i.e., the trace of helicity-2 component $\widetilde{h}_{\mu\nu}$:
 \begin{align}
 \widetilde{h}_{\mu\nu}^{\text{s}} =& - \sqrt{16\pi}\, \Phi\, \eta_{\mu\nu} \,\text{, where} \label{eqn:Cla:ScalarMassiveGraviton}\\
 \Phi =& \frac{\Lambda M_p}{\sqrt{4\pi}m^2} \cos(mt) \,.  \label{eqn:Cla:Phi}
 \end{align}
  We chose numerical prefactors such that  $\Phi$ is canonically normalized.\footnote
 {
 	Canonical normalization means that the term $\partial_\mu \Phi \partial^\mu \Phi$ has the prefactor $\frac{1}{2}$ in the linearized Lagrangian \eqref{massiveTH}.
 } 
   From equation \eqref{eqn:Cla:Chi} it is clear that $\Phi$ also describes $\chi$:
 \begin{align}
 	\chi = 8 \sqrt{4\pi}\, \Phi \,. \label{eqn:Cla:ChiFromPhi}
 \end{align}
 Thus, the field $\Phi$ represents the scalar degree of freedom which simultaneously resides both in the trace of the helicity-2 component $\widetilde{h}_{\mu\nu}$ and in the helicity-0 component $\chi$. 
 
 Clearly, choosing $\Phi$ as in \eqref{eqn:Cla:Phi} amounts to a particular choice of gauge. We selected it for simplicity: It enables us to study linearized de Sitter in terms of only one scalar degree of freedom. We have this gauge freedom at our disposal since the Lagrangian \eqref{massiveTH} is manifestly gauge-invariant. Had we chosen a different gauge, the analysis would generically become more complicated as a single scalar degree of freedom would no longer suffice to describe the de Sitter metric.

\subsection{Definition of the quantum operator and state}

\subsubsection*{Quantum description in terms of a free scalar field}
Our next goal is to understand the classical solution \eqref{eqn:Cla:ScalarMassiveGraviton} as the expectation value of a 
field operator over an appropriate quantum state.  From now on we will set 
$\hbar = 1$.
As said above, we shall focus on the scalar field $\Phi$ which determines both $\widetilde{h}$ and $\chi$.\footnote
{
	This means that we lift the classical split of $\widetilde{h}_{\mu\nu}$ in its trace and traceless components to the quantum level, i.e., we define the quantum state of $\widetilde{h}_{\mu\nu}$ by defining a quantum state for each symmetry component. Subsequently, we can view the full graviton state $\ket{N_{\tilde{h}}}$ as tensor product of them: $\ket{N_{\tilde{h}}} = \ket{N} \otimes_i |0_{h_i}\rangle$, where $|0_{h_i}\rangle$ are the vacua for the other components.
} 
Thus, our task simplifies to defining a free massive scalar field operator  $\hat{\Phi}$ and a corresponding quantum coherent state $\ket{N}$ such that
\begin{align}
\bra{N}\hat{\Phi}\ket{N} = v \cos(mt) \,,
\end{align} 
with $v := \frac{\Lambda M_p}{\sqrt{4\pi}m^2}$ according to \eqref{eqn:Cla:Phi}.

We start from the Lagrangian of a free massive scalar field:
\begin{align}
\hat{\mathcal{L}} = \frac{1}{2}\left((\partial_\mu \hat{\Phi})^2 - m^2 \hat{\Phi}^2\right) \,.
\end{align}
Since there is no interaction, we can expand the full Heisenberg operator $\hat{\Phi}$ in creation and annihilation operators:
\begin{align}
\hat{\Phi} = \intMode \left(\hat{a}_{\vec{k}}e^{-ikx} + \hat{a}_{\vec{k}}^\dagger e^{ikx}\right) \,, \label{eqn:Exp:ModeExpansion}
\end{align}
with the standard commutation relations $[\hat{a}_{\vec{k}},\hat{a}_{\vec{k}'}^\dagger] = \delta^{(3)}(\vec{k} - \vec{k}')$. For later convenience, we also define the operator $\hat{\tilde{a}}_{\vec{k}} := ((2\pi)^3/V)^{1/2}\, \hat{a}_{\vec{k}}$, whose commutator is normalized to 1.

As explained before, it is crucial for the validity of our approach that we consider a free field since only in that case, we can use the Fock space of the free operators $\hat{a}_{\vec{k}}$, $ \hat{a}_{\vec{k}}^\dagger$ to construct the coherent state $\ket{N}$. In the presence of interaction, we would have to use different operators with a modified dispersion relation. Those are, however, generically impossible to define. It is the approximation \eqref{eqn:Cla:MassiveHelicity2Solution} of the classical solution which enables us to use a free field. The
self-consistency of using this free massive scalar field  justifies our choice of classical approximation. At this point, it becomes again clear why it is not convenient to use a free massless field for the quantum description of linearized de Sitter \eqref{eqn:Cla:MasslessSolution}. This is the case since the expectation value of the scalar field \eqref{eqn:Cla:Phi} diverges for $m\rightarrow 0$. Equivalently, the field operator \eqref{eqn:Exp:ModeExpansion} becomes time-independent in the mode of zero momentum.\footnote
{
	 Note that the limit $m\rightarrow 0$ was used in \cite{Dvali:2015Classical} to obtain a coherent state description of Minkowski space  in terms of zero energy gravitons. This connection once again shows the fundamental difference between the two space-times:  The quantum  constituents of de Sitter must carry non-zero energies.
}

Finally, we need to determine the quantum state $\ket{N}$ of the scalar field $\hat{\Phi}$. In full analogy to the anharmonic oscillator, we choose it as coherent state of zero momentum quanta:
\begin{align}
\ket{N} = &\ \text{e}^{-\frac{1}{2}N} \sum_{n=0}^\infty \frac{N^{\frac{n}{2}}}{\sqrt{n!}} \ket{n}  \,\text{, with}  \label{eqn:Exp:QuantumState}\\
N =&\ \frac{ V \Lambda^2 M_p^2}{8 \pi m^3} \,.\label{eqn:Exp:N}
\end{align}
In this formula, $\ket{n} = (\hat{\tilde{a}}_{\vec{0}}^\dagger)^n (n!)^{-\frac{1}{2}} \ket{0}$ are normalized number eigenstates of $n$ quanta with zero momentum. Using the fact that coherent states are eigenvectors of the annihilation operator, we conclude that the state $\ket{N}$ indeed yields the correct expectation value:
\begin{align}
\bra{N}\hat{\Phi}\ket{N} = \frac{1}{\sqrt{ 2m}}\left(\sqrt{\frac{N}{V}}\text{e}^{-imt} + \sqrt{\frac{N}{V}}\text{e}^{imt}\right) = v \cos(mt) \,, \label{eqn:Cla:ExpectationValue}
\end{align}
where we plugged in the mode expansion \eqref{eqn:Exp:ModeExpansion}. Thus, we have found a quantum description of the classical metric \eqref{eqn:Cla:ScalarMassiveGraviton} in terms of the operator \eqref{eqn:Exp:ModeExpansion} and the state \eqref{eqn:Exp:QuantumState}.

\subsubsection*{Construction of the coherent de Sitter state}
For the sake of completeness, we will outline two procedures to construct the coherent state \eqref{eqn:Exp:QuantumState}. The first one is to look for a state which maximizes the classical expectation value, i.e.,
\begin{align}
\frac{\left|\braket{N|\hat{\Phi}|N}\right|}{\braket{N|N}} \,.
\end{align}
This condition amounts to realizing the classical result with a minimal quantum input. Since coherent states are eigenstates of the annihilation operator, this procedure leads to \eqref{eqn:Exp:QuantumState}.

As second justification for the use of coherent states would be to follow \cite{Dvali:2015Classical}. There, the idea is to consider a shift of the state $\Phi = 0$, which corresponds to the classical vacuum, to a different value $\Phi = v$. In the limit $m \rightarrow 0$, such a shift
$\Phi \rightarrow \Phi + v$ by a constant  $v$ would correspond to a symmetry. The corresponding shift-generator  $\hat{Q}$, which would represent a conserved charge for $m \rightarrow 0$, has the form:
\begin{align}
\hat{Q}(t) = \int \diff^3\vec{x}\, \partial_t \hat{\Phi} = -iq \left( \text{e}^{-imt}\hat{\tilde{a}}_{\vec{0}} - \text{e}^{imt}\hat{\tilde{a}}_{\vec{0}}^\dagger\right),
\end{align}
with $q = \sqrt{\frac{m V}{2}}$. Using this generator, we can obtain the coherent state $\ket{N}$ by shifting the vacuum state $\ket{0}$ corresponding to $\Phi = 0$:
\begin{align}
\ket{N} := \exp\left\{-iv\hat{Q}(t=0)\right\}\ket{0} = \exp\left\{\sqrt{N} \left(\hat{\tilde{a}}_{\vec{0}} - \hat{\tilde{a}}_{\vec{0}}^\dagger\right) \right\} \ket{0}\,,
\end{align}
with $\sqrt{N} = vq$, i.e., $N =  \frac{ V \Lambda^2 M_p^2}{8 \pi m^3}$ as in \eqref{eqn:Exp:N}. Note that our states must be time-independent since we are working in the Heisenberg picture. Thus, we have evaluated the charge operator at a fixed time. We could have  introduced a constant phase in this way, but for simplicity we set it to $0$. Using the Baker-Campbell-Hausdorff formula, we finally obtain \eqref{eqn:Exp:QuantumState}.

\subsection{De Sitter as coherent state} \label{sec:ResolutionDiscussion}

\subsubsection*{Energy, number and coupling of the gravitons}
 In \eqref{eqn:Cla:ApproximateMass}, we already observed that we need $m\approx \sqrt{\Lambda}$ for the validity of our approximation of 
the linearized de Sitter metric by an oscillating solution of  
massive gravity.  The same equality is imposed upon us 
by matching the energies of the cosmological constant and the graviton 
gas.  Indeed, in  the quantum description, the expectation value of energy
per volume $V$ is given by a product of the graviton rest mass and their average occupation number in a coherent state: 
 \begin{align}
E_{\text{quant}} = m N = \frac{ \Lambda^2 M_p^2 V}{8\pi m^2} \,. \label{eqn:Exp:TotalEnergy}
\end{align}

In order to be consistent with the known classical energy density associated to the cosmological constant, $E_{\text{class}} = \frac{\Lambda M_p^2 V}{8\pi}$, we set\footnote
{
	We remark that we deal with two different notions of energy. The classical energy is associated to the source $\Lambda$. The energy of the massive gravitons originates from the gravitational field. We could compare this to the situation for a shell of mass $M$: In that case, $M$ is the energy of the source whereas $\frac{M^2}{2M_p^2 R}$ is the energy of the gravitational field.
} 
\begin{align}
m = \sqrt{\Lambda} \,.
\label{Lmass}
\end{align}
While the time evolution of the field $\Phi$ is independent of the graviton mass to leading order, only one value of $m$ correctly reproduces the energy density. This finding matches our initial intuition. In understanding de Sitter as multi-particle quantum state, we started from the observation that the classical solution only contains one energy scale, namely $\sqrt{\Lambda}$. In any quantum description, we therefore expect that the frequency of the constituents should be set by this natural scale. That this is indeed the case is a consistency check for our approach.

In total, we have obtained a description of the de Sitter space-time as  of a  coherent state of gravitons of the mean occupation number 
\begin{align}
	N =  \frac{ V \sqrt{\Lambda} M_p^2}{8 \pi } \label{eqn:Exp:TotalNumber}
\end{align}
and frequency $ m = \sqrt{\Lambda}$. For the mean occupation number in one Hubble volume, $V\propto \Lambda ^{-3/2}$, we get
\begin{align}
	N = \frac{M_p^2}{\Lambda} \label{eqn:Exp:NumberInHubble}
\end{align}
gravitons, in accordance with \cite{Dvali:2013Compositness}.\footnote
{
	 Note that in any case we need to work with volumes $V\lesssim \Lambda ^{-3/2}$ for the validity of our first order approximation.
}
 
Let us now estimate the quantum coupling of the coherent state  gravitons. Despite the fact that the classical solution was obtained in the approximation of ignoring the self-coupling of gravitons, the strength of their coupling is universally fixed by general covariance. To first nonlinear order, the graviton self-coupling can be estimated by taking into account the coupling of gravitons to their own energy-momentum tensor $T_{\mu\nu}(\tilde{h})$ evaluated to bilinear order in 
 $\tilde{h}$.  Even without presenting the explicit long
expression for $T_{\mu\nu}(\tilde{h})$, it is clear that for coherent state gravitons of energy $m$, the strength of the effective four-point coupling is given  by 
\begin{equation}
\alpha = {m^2 \over M_P^2}\,.  
\label{couplingG}
\end{equation}
 Taking into account the relation between graviton frequency 
 and cosmological constant  \eqref{Lmass}  as well as formula \eqref{eqn:Exp:NumberInHubble} for $N$, we obtain the following relation between the coupling and the occupation number of the coherent state  gravitons:
\begin{align}
	\alpha = \frac{\Lambda}{M_p^2} = \frac{1}{N} \,. \label{eqn:Exp:GravitationalCoupling}
\end{align}
This expression reveals a remarkable property of de Sitter understood as the coherent
state:  the collective interaction of the constituent gravitons  is {\it near critical}, $\alpha N = 1$. 
This fact immediately demonstrates the consistency between equation 
\eqref{tclassQ} and equations \eqref{dbreak}, \eqref{dbreakM} and \eqref{eqn:Cla:classicalBreakTime}: The classical break-time is
equal to the de Sitter Hubble radius because the collective coupling $\alpha N$ is of order one. 

\subsubsection*{Interpretation of the graviton mass}

Let us summarize our approach. On the mathematical level, we consider a theory of free massive gravity, in which we fix a coherent state of $N$ gravitons at some initial time. After that, it evolves according to its free time evolution. During the whole regime of validity of the first order approximation, i.e., for $t<t_{\text{cl}}$ (see \eqref{eqn:Cla:classicalBreakTime}), the expectation value over this state reproduces the classical de Sitter solution
of a massless theory.   

Consequently, the following physical picture emerges: At the fundamental level, we deal with a theory of massless gravity with the constant source $\Lambda$. It leads to the formation of a multi-particle state which 
represents a quantum description of the de Sitter metric. The characteristic features of this state are: \\
1) the gravitons are off-shell due to the collective coupling;\footnote
{
	This off-shellness is real in the sense that a detector would measure a particle of non-zero energy and zero momentum.
} and
\\
2) the strength of this collective interaction is critical: $\alpha N =1$. 

Thus, this state is not accessible within the standard perturbation theory, i.e., by expansion in operators $\hat{a}_{\text{free}}^{\dagger}$, $\hat{a}_{\text{free}}$ of free massless gravitons.  However, 
following \cite{Dvali:2013Compositness}, we propose that it is possible to "integrate out" the effect of the cosmological constant source
as well as of collective  interaction and replace it by an effective graviton mass. In other words, for a short time we can model the effect of collective interaction -- putting massless gravitons off-shell -- by means of a mass term in a free theory. Hence, we can approximate the interacting massless gravitons by free massive gravitons, described by the Fock space of $\hat{a}^{\dagger}$, $\hat{a}$.

The first evidence which supports the validity of such a modeling is that the state of free massive gravitons reproduces the correct classical expectation value. Of course, this argument does not suffice, since one can realize a given expectation value in a multitude of ways. Therefore, we will collect further evidence by demonstrating that our framework is constructed such that it automatically reproduces all semi-classical results in de Sitter. We shall accomplish this by coupling the "constituent" gravitons of the coherent state to external quantum particles via the universal gravitational coupling and studying quantum processes due to this interaction.\footnote
{
	In marked difference to other theories, the choice of coupling is unique. This might explain why the corpuscular approach is particularly suited for gravity.
} 
After making sure that these quantum processes correctly account for the known phenomena, we use the example of particle production to show that they also lead to an inevitable loss of coherence and a finite quantum break-time of de Sitter space.

\section{Uncovering the quantum origin of classical evolution} 
\label{sec:SemiClassicalLimit}

\subsection{The semi-classical limit in the coherent state picture} 

\subsubsection*{The standard semi-classical treatment}

  At this point, we have obtained a consistent coherent state description of linearized de Sitter, which  -- for a short enough time interval -- reproduces the classical metric  (\ref{eqn:Cla:MasslessSolution}) as expectation value of the graviton field. The resolution of a background metric in form of a quantum state allows us to achieve the following goals: \\
  1) to  understand standard semi-classical processes -- such as 
 the propagation of a probe particle in a background metric and 
 particle creation by a time-dependent classical metric -- in the language of underlying {\it fully quantum} dynamics; and\\  
  2) to identify new corrections originating from this quantum dynamics which the standard semi-classical treatment is unable to capture. 
  
  In this section, we shall deal with the first point and establish how the effective semi-classical evolution emerges as limiting case of our quantum description. Our starting point is the  quantized version of the Lagrangian \eqref{massiveTH} in which both fields  $h^{\mu\nu}$  and $\Psi$ should be replaced by quantum operators.  To be more precise, since we are focusing on the scalar component of the graviton \eqref{eqn:Cla:ScalarMassiveGraviton}, the effective quantum Lagrangian we 
  work with is,   
   \begin{align}
\hat{\mathcal{L}} = \frac{1}{2}\left((\partial_\mu \hat{\Phi})^2 - m^2 \hat{\Phi}^2   +  (\partial_\mu \hat{\Psi})^2 - m_{\Psi}^2 \hat{\Psi}^2 
\right) +  { \hat{\Phi} \over M_P} \hat{T}^{\mu}_{\mu} (\hat{\Psi})\,.
\label{PhiPsi}
\end{align}
 For simplicity, as external field we use a massive scalar $\hat{\Psi}$ with the stress-energy tensor $\hat{T}^{\mu\nu} (\hat{\Psi}) \equiv -\sqrt{16\pi} \left( \partial^\mu \hat{\Psi} \partial^\nu \hat{\Psi} - \frac{1}{2}\partial^\alpha \hat{\Psi} \partial_\alpha \hat{\Psi} \eta^{\mu\nu} + \frac{1}{2}m_{\Psi}^2 \hat{\Psi}^2 \eta^{\mu\nu}\right)$. In order to simplify notations, we chose the unconventional normalization $-\sqrt{16\pi}$ to account for the corresponding factor in \eqref{eqn:Cla:ScalarMassiveGraviton}.

 In general, the semi-classical treatment corresponds to quantizing weak field excitations on top of a fixed classical background metric, i.e., ignoring any back reaction to the metric from the creation and propagation of quantum particles. In our model, this amounts to quantizing the $\hat{\Psi}$-field in an effective de Sitter space-time created by the classical $\Phi$-field. Thus, we can derive the standard semi-classical evolution of a probe particle $\hat{\Psi}$ in the background classical metric from the effective Lagrangian
   \begin{align}
\hat{\mathcal{L}}_{\hat{\Psi}}^{(\text{eff})} = \frac{1}{2}\left( (\partial_\mu \hat{\Psi})^2 - m_{\Psi}^2 \hat{\Psi}^2 
\right) +  {\Phi_{\text{cl}} \over M_P} \hat{T}^{\mu}_{\mu} (\hat{\Psi})\,,
\label{EFF}
\end{align}
 which can be obtained from \eqref{PhiPsi} when we replace the quantum field $\hat{\Phi}$ by the classical solution, $\hat{\Phi} \rightarrow \Phi_{\text{cl}}$. Here $\Phi_{\text{cl}}$ is  given by \eqref{eqn:Cla:Phi}.  In such a treatment, the only relevant asymptotic quantum states are the initial states $\ket{\text{i}_\Psi}$ and the final states $\ket{\text{f}_\Psi}$ of the $\hat{\Psi}$-field since the background metric is a $c$-number. 

The fact that we are treating $\Phi_{\text{cl}}$ as a fixed classical background means that we are taking the limit: 
\begin{equation} 
\Lambda = m^2 = {\rm fixed},~~   M_P \rightarrow \infty \, . 
\label{Classlimit} 
\end{equation}
In that case, any back reaction from the dynamics of $\hat{\Psi}$ can be  
 ignored and we can treat $\Phi_{\text{cl}}$ as an {\it eternal}  classical background. In the quantum picture, this corresponds to taking 
 the limit
 \begin{equation} 
 \Lambda = m^2 =  {\rm fixed},~~   N \rightarrow \infty \,,
 \label{QuantLimit} 
 \end{equation}
 i.e., to using a coherent state with infinite mean occupation number. It is important to note that we keep $\hbar$ fixed  in both limits. For convenience, we have set $\hbar = 1$.

\subsubsection*{The fully quantum picture}
 As we have seen, the replacement $\hat{\Phi} \rightarrow \Phi_{\text{cl}}$ suffices to obtain the semi-classical limit. However, we are asking for more.  In our theory, this approximation must emerge as a result of fully quantum interactions between the metric-quanta and external particles $\hat{\Psi}$. This means that we would like to  understand this replacement not as an external prescription, but as a result of taking the limit \eqref{QuantLimit}  in the full quantum evolution.  
    
   In order to achieve this, let us first  describe the evolution of a 
   $\hat{\Psi}$-field in the effective semi-classical theory \eqref{EFF}  in the language of an $S$-matrix evolution operator. The non-trivial quantum evolution is due to the last term, which represents the off-diagonal part of the Hamiltonian density:
   \begin{equation}  
   \hat{{\mathcal{H}}}^{(\text{eff})}_{\text{int}} (x)  = {\Phi_{\text{cl}} \over M_P} \hat{T}^{\mu}_{\mu}(\hat{\Psi}) \, .
  \label{intHeff}
  \end{equation} 
 We can derive the quantum evolution in a weak-field perturbation theory in the expansion parameter $\Phi_{\text{cl}} / M_P \ll 1$. To first order in this expansion, we define the effective $S$-matrix evolution operator 
 \begin{equation}
  \hat{{\mathcal{S}}}^{(\text{eff})}  = -i 
  \int \diff^4 x \, \mathbf{T}\left\{\hat{{\mathcal{H}}}^{(\text{eff})}_{\text{int}} (x)  \right\} \,.
  \label{Seff}
  \end{equation} 
 The quantum evolution of $\hat{\Psi}$ is then described by the 
matrix elements between different initial and final states:
  \begin{align}
\mathcal{A} = \braket{\text{f}_{\Psi}|\hat{{\mathcal{S}}}^{(\text{eff})}|\text{i}_{\Psi}}  \,.
\label{matrixeff}
\end{align}
 Of course, since the effective Hamiltonian is explicitly time-dependent,
 the evolution described by the effective $S$-matrix is in general non-unitary. This leads to subtleties in defining the relevant 
 initial and final $S$-matrix states on such a time-dependent background.
 This complication is completely standard and  is a consequence of taking the zero back reaction limit. 
 
 Our immediate goal is not to enter in these well-known issues, but rather to understand the effective semi-classical evolution as the limit of the underlying fully quantum one. For this it is enough to recall that the coherent state $\ket{N}$ is defined in such a way that it reproduces the classical metric in form of the expectation value:
\begin{equation}
	\Phi_{\text{cl}} = \braket{N|\hat{\Phi}|N} \,.
\end{equation} 
 Consequently, the effective semi-classical $S$-matrix operator can be written as
\begin{align}
	\braket{\text{f}_{\Psi}|\hat{{\mathcal{S}}}^{(\text{eff})}|\text{i}_{\Psi}}   
=\left(\bra{\text{f}_{\Psi}}\otimes\bra{N}\right)\ \hat{{\mathcal{S}}}\ \left(\ket{N} \otimes \ket{\text{i}_{\Psi}}\right)
	 \,,
\label{matrixQ}
\end{align}
where $\hat{{\mathcal{S}}}$ is the full quantum $S$-matrix evolution operator 
 \begin{equation}
  \hat{{\mathcal{S}}}  = -i 
  \int \diff^4 x \, \mathbf{T}\left\{\hat{{\mathcal{H}}}_{\text{int}} (x)  \right\} \label{eqn:fullSMatrix}
  \end{equation} 
  defined by the full quantum interaction Hamiltonian
 \begin{equation}  
   \hat{{\mathcal{H}}}_{\text{int}} (x)  = {\hat{\Phi}\over M_P} \hat{T}^{\mu}_{\mu} (\hat{\Psi}) \,.
  \label{intH}
  \end{equation} 
This means that the initial state in the fully quantum picture does not only consist of the external particles $\ket{\text{i}_\Psi}$. Instead, we use the coherent state (describing de Sitter)  with external particles on top of it: $ \ket{N} \otimes \ket{\text{i}_\Psi}$. Likewise, the final state is $ \ket{N} \otimes \ket{\text{f}_\Psi}$.
  
    Of course, the true quantum evolution inevitably implies transitions to final graviton states $\ket{N'}$ which differ from the initial coherent state $\ket{N}$ and in general are not even coherent.  Such transitions  are not equivalent to simply replacing  the graviton field  by its expectation value and lead to departures from semi-classicality. 
 Therefore, equation \eqref{matrixQ} makes the quantum meaning of the semi-classical limit apparent: It corresponds to setting $\ket{N'}=\ket{N}$, i.e., ignoring any back reaction to the graviton state. Thus, the semi-classical $S$-matrix elements, which reproduce the motion   of an external $\hat{\Psi}$-particle in the classical metric, are the subset of fully quantum $S$-matrix elements in which the quantum field  $\hat{\Phi}$ is taken in the same initial and final state $\ket{N}$. 
   
   Notice that this selection of matrix elements is automatic in the limit \eqref{QuantLimit}, due to standard properties of coherent states. This is true since the overlap of coherent states scales as
   \begin{equation}
   	\braket{N+\Delta N|N} = \exp\left(\frac{-\Delta N^2}{8 N}\right) \,,
   \end{equation} 
   so that $\braket{N+\Delta N|N} \approx \braket{N|N} =1$ for $N\rightarrow \infty$. In that case, we can set $\ket{N'} \approx \ket{N}$. This consistently explains why this limit corresponds to a zero back reaction. 
 A similar argument holds for transitions from the initial coherent state 
  to non-coherent ones.
   
Finally, we note that establishing the connection between the semi-classical 
and the quantum $S$-matrix evolutions  sheds new light on the 
standard difficulties of defining in- and out-states of the semi-classical $S$-matrix in a time-dependent external metric, such as de Sitter. The reason is the {\it eternal} nature of the background metric. 
  As we have seen, in the quantum language this {\it eternity} 
 translates to the approximation in which the 
 initial and final states of gravitons can be taken as 
 the same {\it undisturbed} coherent state $\ket{N}$. But for finite $N$,
 this approximation is good only  for a finite time:  For finite $N$, the coherent state cannot be eternal! As we shall see, precisely because of back reaction to it, the coherent state has a characteristic lifetime, which defines the quantum break-time of the system.  This time scales as $N$. Consequently, the coherent state can be treated as truly {\it eternal} only in the limit \eqref{QuantLimit}, i.e., for infinite $N$ and zero coupling.  This makes the whole story self-consistent, at least at the level of the approximate toy model which we posses. Despite its simplicity, this model allows us to capture the key essence of the semi-classical problem as well as of its quantum resolution.  In short, we do not need to worry about defining final $S$-matrix states in the light of the fact that the coherent state $\ket{N}$ itself has a finite lifetime. For processes which happen on time-scales shorter than this life-time, an effective  $S$-matrix evolution can be applied as a valid approximation.

\subsection{Redshift as induced graviton emission}

 Establishing the connection between the semi-classical and fully-quantum 
$S$-matrix descriptions gives us an opportunity to understand the underlying 
quantum meaning of classical evolution. As a concrete example, we will discuss the redshift which a probe particle experiences in a classical de Sitter metric. In the semi-classical $S$-matrix description, this process corresponds to an initial state $\ket{\text{i}_{\Psi}} = \hat{b}^\dagger_{\vec{p}}\ket{0}$ of 4-momentum $p=(p_0, \vec{p})$ which has a higher energy than the final state $\ket{\text{f}_{\Psi}} = \hat{b}^\dagger_{\vec{p}'}\ket{0}$  of 4-momentum $p'=(p'_0, \vec{p}')$. Here we denote the creation operators of $\hat{\Psi}$ by $\hat{b}_{\vec{p}}^\dagger$. The corresponding amplitude is
 \begin{align}
\mathcal{A} = \braket{0|\hat{b}_{\vec{p}'} \hat{{\mathcal{S}}}^{(\text{eff})}\hat{b}^\dagger_{\vec{p}}|0}  \,.
\end{align}
A complication arises since the external particle does not propagate on a Minkowski background, but in a time-dependent de Sitter metric so that true non-interacting out-states, which would be required for the $S$-matrix calculation, do not exist.\footnote
{
   As the de Sitter metric is only invariant under spatial but not under time translations, solely the momentum of the external particle is conserved, unlike its energy. This means that the dispersion relation of a $\hat{\Psi}$-particle is not Poincar\'{e}-invariant but depends on time because also asymptotically, it never stops interacting with the effective background metric. Therefore, non-interacting asymptotic states do not exist. A strategy to overcome this difficulty could be to approximate the initial and final dispersion relations as different but constant. Since the time-dependent change of the dispersion relation scales with the Hubble energy $m$, we expect this to be possible if we restrict ourselves to $p_0, p'_0 \gg m$. 
}
This problem does not concern us since it only pertains to the semi-classical treatment. What we are only interested in is mapping the fully-quantum calculation to the semi-classical one.

By the correspondence \eqref{matrixQ} we discussed before, the fully-quantum amplitude is
 \begin{align}
\mathcal{A} = \braket{N'|\hat{b}_{\vec{p}'} \hat{{\mathcal{S}}}\hat{b}^\dagger_{\vec{p}}|N}  \,.
\end{align}
For the purpose of later generalization, we kept $\ket{N'}$ arbitrary, but  we will soon specialize to $\ket{N'} = \ket{N}$. Plugging in the full $S$-matrix operator \eqref{eqn:fullSMatrix}, we obtain
\begin{align}
\mathcal{A} = \mathcal{K}(p, p') \int \diff^4 x \, \text{e}^{-i(p-p')x} \braket{N'|\hat{\Phi} |N} \,, \label{eqn:Sca:AmplitudeRedshift}
\end{align}
where the gravitational field solely appears in $\braket{N'|\hat{\Phi} |N}$ and the kinematical factor $\mathcal{K}(p, p')$ only depends on the external particles. We do not need its explicit form in the present discussion but write it down for the purpose of later use:
\begin{align}
\mathcal{K}(p, p') =&\, i \frac{\sqrt{16\pi}}{M_p} \zeta_{\Psi}(p)\zeta_{\Psi}(p') \left( p\cdot p' - 2 m_{\Psi}^2 \right)  \label{eqn:Sca:KinematicalFactorRedshift} \,,
\end{align}
with the abbreviation $\zeta_{\Psi}(p) = \left((2\pi)^3 2 p_0\right)^{-1/2}$. 

As already discussed, we see explicitly that we can achieve the semi-classical limit $\hat{\Phi} \rightarrow \Phi_{\text{cl}}$  by setting $\ket{N'} = \ket{N}$. Plugging this in as well as the mode expansion \eqref{eqn:Exp:ModeExpansion}, we obtain
\begin{align}
\mathcal{A} =&\, \frac{(2\pi)^4}{\sqrt{2m}}  \mathcal{K}(p,p') \delta^{(3)}(\vec{p} - \vec{p}')\nonumber\\
&\left(\delta(-p_0+p_0'+m) \sqrt{\frac{N}{V}} + \delta(-p_0 + p_0'-m)\sqrt{\frac{N}{V}}\right) \,. \label{eqn:Sca:RedshiftSMatrix}
\end{align}

\begin{figure}
	\begin{center}
		\includegraphics[width=0.65\textwidth]{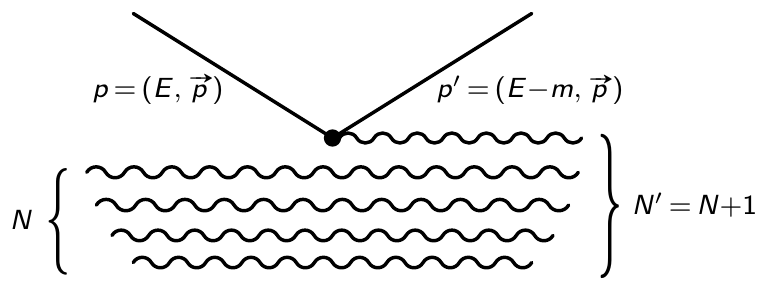}
		\caption{Redshift as stimulated graviton emission: An external particle of initial 4-momentum $p$ deposits a graviton in the background state of $N$ gravitons. The final 4-momentum of the external particle is $p'$.}
		\label{fig:Redshift}
	\end{center}
\end{figure}

This formula makes the quantum dynamics of this process transparent. The external particle emits (contribution $\propto \delta(-p_0+p_0'+m)$) or absorbs (contribution $\propto \delta(-p_0+p_0'-m)$) a background graviton. The emission of a graviton, during which the external particle looses energy, corresponds to redshift whereas the absorption of a graviton leads to an increased energy, i.e., blueshift.\footnote
{
 From our perspective of short time-scales, these two processes are indistinguishable. We expect that the boundary conditions of the expanding de Sitter branch select redshift.
}
In the case of redshift, we furthermore observe that we deal with a process of induced emission which is enhanced by the $N$ already existing gravitons. In the fully-quantum $S$-matrix language, redshift therefore corresponds to the induced emission of a graviton, as already suggested in \cite{Dvali:2013Compositness}. We depict this process in figure \ref{fig:Redshift}.

It is important to note that the final 4-momentum of the emitted graviton is completely fixed in this process of induced emission. Therefore, also the 4-momentum of the external particle is uniquely determined.\footnote
{
	Namely, we have  $p'=(p_0 -m, \vec{p})$. This shows that the dispersion relation of the $\hat{\Psi}$-particle has to change, as we already pointed out. 
}
In this way, our fully quantum computation is able to reproduce the classical redshift.

As already discussed, we deal with a process of induced emission, which is enhanced by the $N$ already existing gravitons. Consequently, we can obtain the amplitude $\mathcal{A}_{\text{spont}}$ of spontaneous emission in the de Sitter background by removing this enhancement: $\mathcal{A}_{\text{spont}} = \mathcal{A}/\sqrt{N}$. This relation is particularly interesting in the semi-classical limit \eqref{QuantLimit}, which corresponds to $N=\infty$. Since the amplitude $\mathcal{A}$ of redshift is finite, we conclude that the coherent state representation of the geometry produces the classical redshift although the amplitude of spontaneous emission $\mathcal{A}_\text{spont}$ is zero in this limit. This is the essential difference between redshift and standard loss of energy by gravitational radiation.\footnote
{
 Of course, as all other processes in our picture, the classical redshift is corrected by quantum $1/N$-effects.
}
 In the corpuscular resolution of de Sitter, the classical process of redshift is therefore fully analogous to a phenomenon of a non-vanishing stimulated emission with zero spontaneous emission. This phenomenon takes place due to the representation of the graviton background as coherent graviton state  whose mean occupation number is infinite in the semi-classical limit \eqref{QuantLimit}.
 	
Heuristically, the process of redshift is analogous to the transitions between energy levels in an atom. In this picture, the initial "atom" $\hat{\Psi}$ emits a graviton under the influence of a coherent state  gravitational "radiation" and gets deexcited to a lower energy state $\hat{\Psi}'$. Clearly, the mass of the atom has to change in the course of this process. 
The difference between the atom in a radiation field and a particle in de Sitter is that the atom possesses different energy levels even without radiation whereas there are no energy levels for a particle in Minkowski.\footnote
{
 This means that the particle is analogous to an atom with degenerate energy levels which only split in the presence of external radiation.
}

\subsection{Dilution of gas as conversion process}

To conclude this section, we want to briefly point to another process, namely the dilution of a gas of massive neutral particles in a de Sitter background, which at the classical level is described by a coherently oscillating 
real scalar field with Hubble friction term:\footnote{
	Note that in the presence of a chemical potential, the story is a bit more involved and will not be considered here.
}  
\begin{equation} 
\ddot{\Psi} \, + \,\sqrt{3\Lambda} \dot{\Psi}  \, + \,m_{\Psi}^2 \Psi \, = \, 0 \,. 
\label{dilution} 
\end{equation} 
We restrict ourselves to the regime where the gas only leads to a small perturbation of the pure de Sitter metric. In the standard classical treatment, one would attribute the dilution to the  Hubble damping given by 
\eqref{dilution}.

\begin{figure}
	\begin{center}
		\includegraphics[width=0.7\textwidth]{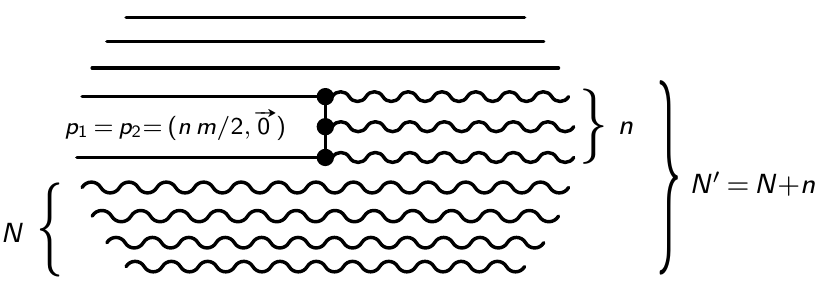}
		\caption{Dilution of gas as conversion of the gas particles: Two gas particles of 4-momentum $p_1$ and $p_2$, which are at rest, annihilate into $n$ gravitons. In this process, both the graviton and the gas state are coherent.}
		\label{fig:Dilution}
	\end{center}
\end{figure}

In our fully-quantum treatment, a different picture emerges. Just like the
de Sitter metric, also the gas has a quantum description as coherent state of $\hat{\Psi}$-particles. The coupling of $\hat{\Psi}$ to gravity makes possible a process of induced decay, which is depicted in figure \ref{fig:Dilution}. For simplicity, we restricted ourselves to $m_\Psi = n\, m/2$, with $n$ integer. In that case, two $\hat{\Psi}$-particles can annihilate into $n$ gravitons. In this manner, the mean number of $\hat{\Psi}$-particles and therefore the density of the gas decreases. The classical condition that the gas only leads to a small perturbation from pure de Sitter amounts to a small back reaction on the quantum level, i.e., to the condition that the change of the graviton state $\ket{N}$ is negligible. This is the case if $m_\Psi N_\Psi \ll m N$, where $N_\Psi$ is the mean occupation number of the coherent $\hat{\Psi}$-state. In summary, at the quantum level, the dilution of gas is caused by a real conversion 
process between $\hat{\Psi}$- and $\hat{\Phi}$-quanta. 

\section{Gibbons-Hawking particle production as quantum decay of the coherent de Sitter state}
\label{sec:ParticleProduction}

\subsection{$S$-matrix approach instead of semi-classical calculation}

\subsubsection*{Relationship to semi-classical approach}
It has been a long-standing prediction by Gibbons and Hawking that particle production occurs in a de Sitter background \cite{Gibbons:1977Cosmological}: An observer should see a thermal spectrum of temperature $T\sim \sqrt{\Lambda}$. In their semi-classical treatment, in which quantum fields are studied on top of the undisturbed classical metric background, particle production arises as a {\it vacuum process}:  Since the vacua of quantum fields depend on time in the evolving de Sitter metric, their early-time vacuum contains particles from the point of view of a late-time observer. As already explained, this semi-classical treatment does not contain a back reaction, i.e., particle production does not change the classical de Sitter metric. 

Having already discussed other processes such as redshift in our quantum description of de Sitter,  our goal is to study particle production in detail. Of course, our treatment is simplified. In particular, we are able to reproduce the classical de Sitter background as the expectation value  over a coherent state only for time-scales shorter than the Hubble time: $t_{\text{cl}} =m^{-1} = 1/\sqrt{\Lambda}$. As explained above,  this is the price we pay for being able to give a meaningful quantum resolution of de Sitter space.  
 So it should not come as a surprise if we will not  be able to reproduce precise features of Gibbons-Hawking particle creation over all times even when we take the semi-classical limit of our calculation. However, we manage to reproduce essentially all the key short-distance features of Gibbons-Hawking radiation, i.e., those features which deal with
 time-scales within the validity of our approximation.  
  
More importantly, the quantum resolution also comes with a great bonus: We are able to capture effects which are not visible in the semi-classical treatment even when the nonlinearities are fully taken into account. It will become clear shortly that this amounts to capturing $1/N$-effects. Despite the fact that we ignore effects of higher order in $\alpha N$, which naively seem much more important, the $1/N$-effects we capture are qualitatively different.  As explained several times, in contrast to $1/N$-effects, the phenomena described by $\alpha N$ are not new and are visible already in semi-classical treatment in form of classical nonlinearities.  

Particle production in our quantum description of the de Sitter metric arises as a Hamiltonian process of scattering and decay of the 
gravitons which compose the coherent state.  Once we take into account the coupling of gravitons to other species  and to each other, inevitably there emerge quantum processes in which part of the coherent state gravitons gets converted into "free" quanta (both in gravitons and in other species). In this context, "free" quanta are those with dispersion relations of particles propagating on top of a classical de Sitter background.  As we are working to linear order in the de Sitter metric, the free quanta will have dispersion relations of Minkowski quanta to the leading order, with the de Sitter metric being a small correction.  As said above, due to the limitations of our approach, it only makes sense to take into account leading order effects.  It is surprising how well this approximation reproduces the qualitative features of Gibbons-Hawking radiation.  

To summarize, Gibbons-Hawking particle production arises in our quantum description of the de Sitter metric in the following way: The self-coupling of gravitons as well as  their coupling to other particle species lead to processes of quantum scattering and decay of the constituent gravitons of de Sitter. The final products of such decays and scatterings contain particles which no longer belong to the coherent state and have dispersion relations of free quanta propagating on a classical de Sitter background. In the usual semi-classical treatment, these processes correspond to the production of Gibbons-Hawking quanta. 

\subsubsection*{Deviations from semi-classicality: $1/N$-effects}
\begin{figure}
	\begin{center}
	\includegraphics[width=0.65\textwidth]{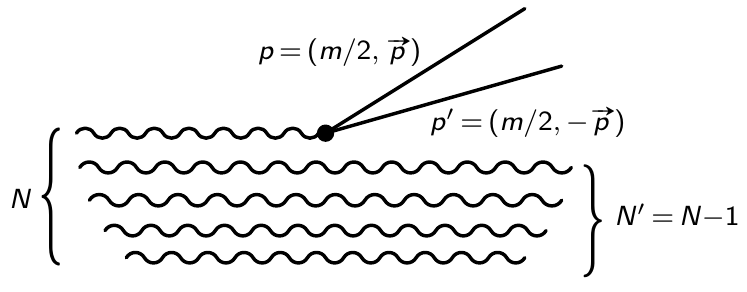}
	\caption{Particle production as graviton decay: One of the $N$ initial gravitons decays and produces 2 external particles of 4-momentum $p$ and $p'$. }
	\label{fig:ParticleProduction}
	\end{center}
\end{figure}

We shall proceed to make more quantitative estimates. For illustrating the point, we limit ourselves to studying the decay processes of background gravitons into external particle species. Due to the universality of graviton coupling, the process of graviton production has a similar rate. The simplest process contributing to particle production via decay is given by a transition from the initial state $\ket{N}$ to a final state which contains two particles other than gravitons. It is depicted in figure \ref{fig:ParticleProduction}. The corresponding amplitude is determined by the one of redshift \eqref{eqn:Sca:AmplitudeRedshift}, after the substitution $p \rightarrow -p$ (where now both and $p$ and $p'$ correspond to outgoing external particles):
\begin{align}
\mathcal{A} =  \braket{N'|\hat{b}_{\vec{p}} \hat{b}_{\vec{p}'} \hat{{\mathcal{S}}}|N} = \mathcal{K}(-p, p') \int \diff^4 x \, e^{i(p+p')x} \braket{N'|\hat{\Phi} |N} \,.\label{eqn:Sca:AmplitudeHelicity2}
	\end{align}

In comparison to the semi-classical treatment, the crucial novelty of our approach is that it uncovers the existence of quantum processes in which the final state $\ket{N'}$ of background gravitons is different from the initial one $\ket{N}$.  These processes correspond to  $1/N$-corrections and therefore are fundamentally invisible 
in the semi-classical picture which, as explained above, in our 
framework is reproduced in $N \rightarrow \infty$ limit. In particular, the final state $\ket{N'}$ obtained as a result of particle production in quantum theory does not even have to be a coherent state. 
 This deviation from coherence gives a gradual departure from classicality and eventually leads to a quantum break-time. 
 
 However, for the case of de Sitter even the transitions among coherent states with different occupation numbers are sufficient for capturing the departure from the classical evolution since classically de Sitter is an eternal state and simply cannot change. Any quantum process  
which changes the characteristics of de Sitter space marks a 
fundamentally new phenomenon not visible in the 
semi-classical theory.  
  Therefore, for illustrating this point we shall consider transitions to a coherent state, 
but with different mean occupation number $N'\neq N$.\footnote
{
	As we explained around equation \eqref{classEF}, we expect that a pure decay process does not lead to decoherence for the anharmonic oscillator because the corresponding matrix element only contains the operator $\hat{a}_{\vec{k}}$. In section \ref{sec:QuantumBreakTime} we will argue, however, that this argument does not apply to our quantum description of de Sitter.
}

Finally, we need to determine the expectation value of the particle number for the final coherent state. In doing so, we want to disentangle two effects: Since a coherent state is not a particle number eigenstate, its particle number fluctuates even without any interaction, i.e., $\braket{N'|N} \neq 0$. Our goal is not to consider this non-Hamiltonian effect. Instead, we want to focus on how the coherent state evolves due to interaction of 
constituent gravitons among themselves and 
with the external particles. Therefore, it is most natural to consider the process which would also be possible if we were to replace the coherent background states by number eigenstates. As we consider the decay of a particle, this leads to $\Delta N = N' - N = -1$.

 Moreover, this choice of $\Delta N$ conserves the expectation value of the energy, as it should. Since the coupling of the coherent state and the produced particles vanishes asymptotically, the only relevant contributions to the Hamiltonian come from the free massive gravitons, $\hat{H}_0^{(\Phi)}$, and from the free external particles, $\hat{H}_0^{(\Psi)}$:
\begin{align}
   	\braket{N|\hat{H}_0^{(\Phi)}|N} = \braket{N'|\hat{H}_0^{(\Phi)}|N'} + \braket{0|\hat{b}_{\vec{p}} \hat{b}_{\vec{p}'} \hat{H}_0^{(\Psi)} \hat{b}^\dagger_{\vec{p}} \hat{b}^\dagger_{\vec{p}'}|0} \,.
\end{align}
This simple argument in terms of the free Hamiltonians only works, however, when gravity couples to an external field, as we assumed in our calculation. If scattering happens due to self-coupling, then the interaction term gives a non-zero contribution to the energy even asymptotically because of graviton interaction in the coherent state. Only if one could calculate this contribution, it would be possible to argue in terms of the energy expectation value. In any case, the precise value of $\Delta N$ does not matter for our conclusions.

\subsubsection*{Calculation of the quantum rate}
Using $\Delta N \ll N$, we obtain the matrix element:
\begin{align}
\braket{N'|\hat{\Phi}|N}= \frac{1}{\sqrt{2 m V}} \left( e^{imt} \sqrt{N'}  + e^{-imt} \sqrt{N} \right)\left(1-\frac{\Delta N^2}{8N}\right)\,, \label{eqn:Exp:ScatteringMatrixElement}
\end{align}
where the $1/N$-correction comes from the overlap of different coherent states. It vanishes in the semi-classical limit $\Delta N =0$. The $S$-matrix element subsequently becomes:
\begin{align}
\mathcal{A} =&\,  \frac{(2\pi)^4}{\sqrt{ 2mV}}  \mathcal{K}(-p, p') \delta^{(3)}(\vec{p} + \vec{p}')\left(1-\frac{\Delta N^2}{8N}\right) \nonumber \\
	&\left(\delta(p_0 + p_0' -m) \sqrt{N'} + \delta(p_0 + p_0' + m)\sqrt{N}\right) \,.  \label{eqn:Sca:ProductionSMatrix}
	\end{align}
After the substitution $p\rightarrow -p$, the matrix element reduces to the result \eqref{eqn:Sca:RedshiftSMatrix} for redshift in the limit $N'=N$, as it should. The amplitude of particle production consists of two parts. The first one describes a process where a graviton leaves the bound state and the second one corresponds to adding a graviton to the bound state. In contrast to the case of redshift, the second process cannot occur because of energy conservation so that the term will be dropped.

As is derived in appendix \ref{sec:AppendixParticleProduction}, the rate of particle production is 
		\begin{align}
			\Gamma =  \frac{2 \sqrt{\frac{m^2}{4} -m_\Psi^2} N}{ M_p^2 m^2 }\left(\frac{m^2}{2} + m_{\Psi}^2 \right)^2 \left(1-\frac{\Delta N^2-4\Delta N}{4N}\right) \,. \label{eqn:Sca:ProductionGamma}
		\end{align} 
		We observe that particle production to first order only takes place for light particles, $m_\Psi < \frac{m}{2}$, as we expect it. In order to simplify the discussion, we specialize to massless external particles ($m_\Psi = 0$):
	\begin{align}
		\Gamma = \frac{N m^3}{4 M_p^2  } (1-\frac{\Delta N^2-4\Delta N}{4N}) = \frac{\Lambda^{2} V}{32\pi} (1-\frac{\Delta N^2-4\Delta N}{4N})\,.
	\end{align}
For dimensional reasons, this result does not come as a surprise. The rate must be proportional to the volume $V$. Since particle production can also be derived in the semi-classical treatment, i.e., when only $\hat{\Psi}$ is quantized but gravity is treated classically, the Planck mass $M_p$ should not appear so that we can only use $\Lambda$ to obtain the correct mass dimension. 
		
In one Hubble volume $V \propto \Lambda^{-1.5}$, this result matches \cite{Dvali:2013Compositness}, where it is obtained using the following heuristic argument: The rate should be the product of the 
number $N$ of particles, the coupling $\alpha$ and the characteristic energy $E$. According to \eqref{eqn:Exp:NumberInHubble} and \eqref{eqn:Exp:GravitationalCoupling}, we have $N\propto M_p^2/\Lambda$, $\alpha \propto \Lambda/M_p^2$ and $E\propto\sqrt{\Lambda}$. This implies $\Gamma = \sqrt{\Lambda}$, the same result we obtained here.

\subsection{Glimpses of Gibbons-Hawking temperature}	

\subsubsection*{Relation to the semi-classical calculation} \label{sec:ProductionDiscussion}

Before we proceed to our main result, the quantum break-time, we want to check to what extent our approach is consistent with the semi-classical result. To this end, we first estimate the power of produced particles. In the semi-classical treatment, the Hubble horizon radiates like a black body of temperature $T\sim \sqrt{\Lambda}$. According to the Stefan-Boltzman law, this yields the emitted power $P \sim T^4 A$, where $A$ is the area of the horizon. Since $A \sim \Lambda^{-1}$, we get the semi-classical power
\begin{align}
	P_{\text{s-c}} \sim \Lambda \,. \label{eqn:Sca:ClassicalPower}
\end{align} 
As a pair of produced particles has the energy $\sqrt{\Lambda}$, our quantum result for the decay rate \eqref{eqn:Sca:ProductionGamma} leads to the consistent result
\begin{align}
	P_{\text{q}} \sim \Lambda \,.
\end{align}
Restoring factors of $\hbar$ for a moment, $P_{\text{q}} \sim \hbar \Lambda$, we note that particle production vanishes in the classical limit $\hbar \rightarrow 0$, as it should.
	
	\subsubsection*{Distribution of the produced particles:  evidence for 
	Gibbons-Hawking thermality}
Having concluded that the total power of produced particles is of the right order of magnitude, we proceed by investigating the distribution of produced particles. To first order, they are not distributed thermally, but all have the same energy $\frac{m}{2}$. As soon as one goes to higher orders so that more than one background graviton participate in the scattering process, this $\delta$-distribution will be smeared out. Furthermore, we expect at least qualitatively that the resulting distribution is thermal. 

\begin{figure}
	\begin{center}
		\includegraphics[width=0.7\textwidth]{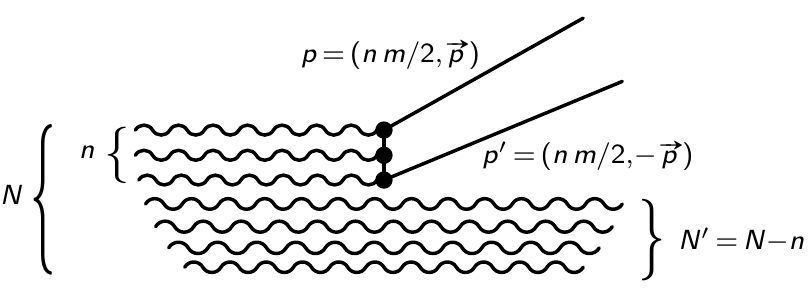}
		\caption{Leading order process for the production of particles with high total energy $E \gg m$. At least $n>E/m$ of the initial $N$ gravitons have to decay to produce the two external particles of 4-momentum $p$ and $p'$.}
		\label{fig:HighEnergy}
	\end{center}
\end{figure}

For example, one of the key features of the Gibbons-Hawking thermal spectrum is the Boltzmann suppression of the production rate of particles of 
energy  $E$ higher  than the de Sitter Hubble parameter $\sqrt{\Lambda}$: 
\begin{equation}
\Gamma \propto  {\rm e}^{-{E \over \sqrt{\Lambda}}}\,.
\label{B-factor} 
\end{equation} 
Our quantum description  of  de Sitter space gives a very interesting
microscopic explanation of this suppression.   The production of particles is a Hamiltonian process in which the background coherent state gravitons get converted into free quanta.  Since the  frequencies of background gravitons are given  by $m = \sqrt{\Lambda}$, the production of states with energies $E \gg m$ requires the annihilation of $n$ background gravitons, with $n > E/m$.  The leading order contribution to this process is schematically expressed in figure \ref{fig:HighEnergy}, which depicts the annihilation of $n$ gravitons into a pair of $\Psi$-quanta with total energy $E = nm$.  The probability of such a process is highly suppressed due to the participation of a large number of soft gravitons in it: Each soft vertex contributes a factor $\alpha = 1/N$ to the rate.  

For estimating this suppression, we can directly use the results of  
\cite{Dvali:20142ToN}, where the multi-graviton transition amplitudes 
have been calculated (see \cite{Addazi:2016BlackHoleFormation} for a related discussion). Accounting for the fact that there are ${N \choose n}$ possibilities to choose the $n$ annihilating gravitons, we obtain the rate: 
	\begin{equation}
	\Gamma \propto   \left ( {1 \over N} \right )^{n} n!\ {N \choose n}\,.   
	\label{rateN3} 
	\end{equation} 
	Using Stirling's formula twice, which is valid for $N\gg 1$ and $N-n\gg1$, we get
	\begin{align}
		\Gamma \approx   \text{e}^{-n} \left(\frac{N}{N-n}\right)^{N-n}\,. \label{rateApproximate}
	\end{align}
	Before we discuss the exponential suppression, we analyze the additional factor $\left(\frac{N}{N-n}\right)^{N-n}$. Defining $l = N/(N-n)$, we can rewrite it as $l^{N/l}$. It is $1$ for $l=1$ and $l \rightarrow \infty$.\footnote
	{
	   The approximation \eqref{rateApproximate} is no longer valid for $l\rightarrow\infty$, i.e., $n=N$, but we can directly read off from \eqref{rateN3} that the additional factor is one: $\Gamma \propto {\rm e}^{- N}$. 	
	} 
	Its maximum is at $l=\text{e}$ and gives $\text{e}^{N/\text{e}}$. At this point, the rate is still exponentially suppressed:
	\begin{align}
		\Gamma_{\text{max}} \propto \text{e}^{-\frac{\text{e} - 2}{\text{e}-1}n} \,.
	\end{align}
	Thus, we conclude that 
	\begin{align}
		\Gamma \propto \text{e}^{-c(n)\, n} \,,
	\end{align}
	with $c(n) \approx 1$ and $c(N)=1$. Since we have $n = E/m$ and $m = \sqrt{\Lambda}$, the above expression reproduces the exponential suppression of the Boltzmann factor (\ref{B-factor}) for the thermal bath. For $n=N$, this correspondence even becomes exact. It is remarkable that the analysis of multi-graviton scattering suffices to obtain a thermal behavior.

\begin{figure}
	\begin{center}
		\includegraphics[width=0.65\textwidth]{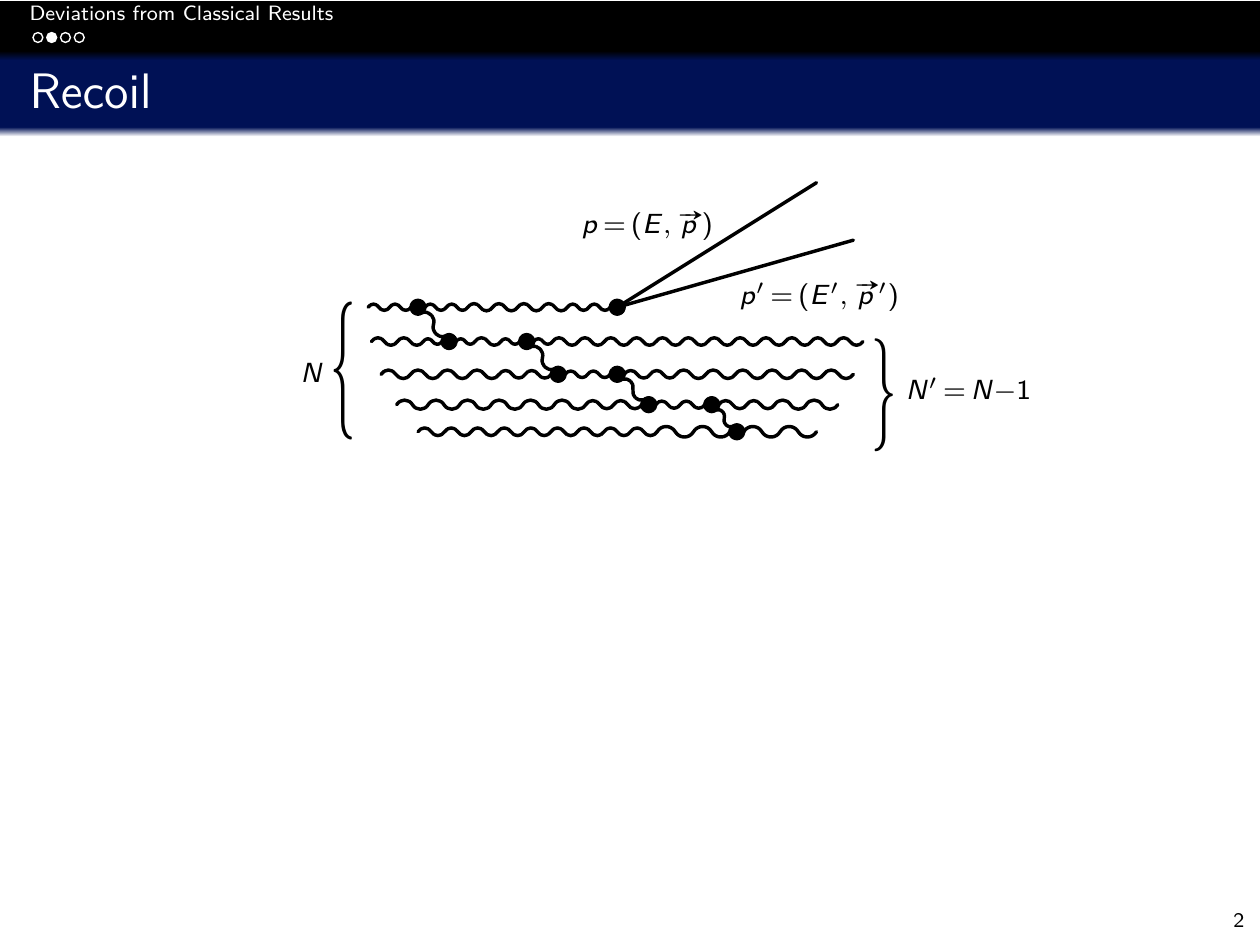}
		\caption{Higher order process of particle production, in which the produced particles recoil against all remaining gravitons. In particular, this allows for produced particles of low energies $E, E'\ll m/2$.}
		\label{fig:Recoil}
	\end{center}
\end{figure}

Having studied produced particles of high energies, we note that smearing out of the energy spectrum also takes place on the infrared side.  Indeed, we can produce arbitrarily soft quanta in processes of decays or annihilations of the background gravitons by redistributing the rest of the energy among the remaining background gravitons in form of a recoil. For example, one of the background gravitons can decay into a pair  of  $\Psi$-particles of energy $E \ll m$ while transferring the energy difference $\Delta E = m - E$  to the remaining $N-1$ gravitons.
 The process is schematically depicted in figure \ref{fig:Recoil}. Unlike 
the case of $E \gg m$, the processes with $ E \ll m$ are 
not exponentially suppressed. 

It would not be very informative to give a more precise estimate of processes with the production of deep-infrared quanta of energy $E \ll m$ since the wavelength of such particles exceeds the size of the de Sitter Hubble patch and thus the time-scale of validity of our model.   
So in the discussion of very low energy particle production, we will be satisfied with qualitative arguments, which clearly show that -- even ignoring classical nonlinearities and formally extrapolating our model for longer times -- the production of very soft modes is possible due to  multi-graviton processes.  

  We thus see that by taking into account multi-graviton contributions to particle production, the $\delta$-distribution gets smeared out.  Its peak is still at $E \sim m$, but  the region  $E \gg m$ is exponentially suppressed by a factor that is strikingly similar to a thermal Boltzmann suppression factor (\ref{B-factor}).  There are a few things remarkable about this result.  First, we see that the thermal-like distribution of the produced particles can result without any need of the notion of temperature in the microscopic theory, but rather as a result of the structure of soft multi-graviton amplitudes.  
Secondly, we have achieved all this within a simplest framework in which the classical part of de Sitter is mapped on a coherent state of gravitons of a linear theory. The quantum effects of de Sitter -- such 
as Gibbons-Hawking radiation -- result from quantum interactions of these 
gravitons. Of course, we do not expect that this simple theory can quantitatively 
capture all the properties of de Sitter, but it already took us surprisingly far. 
	
	\subsubsection*{Consistency check: Gauge invariance}
	
As a final consistency check, we consider a brief example of how the rate of particle production transforms under gauge transformations. To this end, we consider the first order of the transformation from conformal time $t$ to physical time $t'$: $\diff t = a(t) \diff t'$, where $a(t) \propto \cosh(\sqrt{\Lambda/3}\,t)$ is the scale factor. Since $a(0) =1$, we expect this not to have any influence on the decay rate in the semi-classical treatment. Our goal is to show how this gauge invariance arises in our framework. Since the leading order of the gauge transformation changes the metric as $g'_{00} - g_{00} \propto \Lambda t^2$, our first order perturbation generalizes to
\begin{align}
\widetilde{h}'_{00} =& c\, \widetilde{h}_{00} \,, \label{eqn:Sca:GaugeTrafo}
\end{align}
where $c=1$ yields the untransformed value.\footnote
{
	 This corresponds to the infinitesimal transformation $\xi^\mu \propto \delta_0^\mu c\,\Lambda t^3$. It destroys de Donder gauge, but the resulting metric still is a solution of the ungauged equations of motion.
}
For the quantum description in terms of interacting gravitons, $c\neq 1$ means that we have to introduce two different scalar function for $h_{00}$ and the spatial diagonal perturbations.
	
Generalizing our calculation of the first order matrix element, we observe that only the kinematical factor \eqref{eqn:Sca:KinematicalFactorRedshift} changes:
\begin{align}
	\mathcal{K}'(-p,p') =&\, i \frac{\sqrt{16\pi}}{M_p} \zeta_{\Psi}(p)\zeta_{\Psi}(p') \nonumber\\
	& \left(  -(-\frac{1}{2}c + \frac{3}{2})p_0 p'_0 + (\frac{1}{2}c + \frac{1}{2} )\vec{p}\vec{p}'  - (\frac{1}{2}c + \frac{3}{2})  m_{\Psi}^2 \right) \,. \label{eqn:Sca:GaugetransformedKinematicalFactor}
\end{align}
Plugging in the kinematics of a decay, we conclude that the kinematical factor and thus also the rate are indeed gauge-invariant:\footnote
{
	This result also determines how the decay rate changes under rescaling of the spatial components,
	\begin{align*}
		\widetilde{h}'_{11}=&\widetilde{h}'_{22}=\widetilde{h}'_{33} = b\, \widetilde{h}_{11} \,.
	\end{align*}
	Since we have shown that a change of $h_{00}$ has no effect, the spatial transformation is equivalent to a rescaling of all perturbations, $\widetilde{h}'_{\mu\nu} = b\, \widetilde{h}_{\mu\nu}$. This can be absorbed by the redefinition
	\begin{align*}
		\Lambda' = b\, \Lambda \,.
	\end{align*}
	Both in the classical and the quantum description, all results will change according to a modified cosmological constant.
}
\begin{align}
    \left|\mathcal{K}'(-p,p')\right|^2 = \left(\frac{m^2}{2} + m_{\Psi}^2\right)^2 \,.
\end{align}
	
\subsection{Quantum break-time} \label{sec:QuantumBreakTime}

\subsubsection*{Decoherence because of decay}	
For the anharmonic oscillator \eqref{class}, we observed that self-interaction, $\phi + \phi \rightarrow \phi + \phi$, leads to decoherence of the initial coherent state. After a significant number of quanta has left the initial state, the expectation value of the quantum-evolved state will deviate significantly from the classical solution. The corresponding time-scale is the quantum break-time $t_{\text{q}}$. When we added a decay channel (see Lagrangian \eqref{classNEW}), $\phi \rightarrow \psi + \psi$, we argued that
 it changes the classical evolution. However, as long as no back reaction from one-particle decay, such as rescattering or recoil, is taken into account, the coherence of the state of $\phi$-quanta is not lost.  Correspondingly, 
the evolution can be traced by an effective classical equation
(\ref{classEF})
 with an additional friction term.  Of course, in reality any quantum decay process 
will inevitably be accompanied by rescattering and the loss of coherence, 
which will become important sooner or later.  

 In the case of de Sitter, the story is even more dramatic. 
Even if hypothetically we were allowed to maintain only processes which preserve the coherence of the state of background gravitons, this would 
anyway inevitably lead us to a quantum evolution which has no counterpart in the semi-classical picture: Semi-classically, de Sitter is an eternal state with no clock. Therefore, even a  "clean" transition between the two different coherent states  $\ket{N}$ and $\ket{N'}$ gives an intrinsically quantum evolution of the state which cannot be matched by anything in the classical theory. 

As we have seen, it is moreover impossible to maintain only processes which 
preserve the coherence of the state. In particular, the decay process which we calculated explicitly also leads to decoherence. This is true since 
even if we did not include the self-interaction of gravitons, the exchange by virtual $\hat{\Psi}$-particles will lead to their rescattering and subsequent decoherence. In other words, decay and rescattering go hand in hand due to the structure of spin-2 coupling.

\subsubsection*{Estimation of the time-scale}
In summary, we have found a crucial difference to the semi-classical treatment: In our quantum description of de Sitter, the back reaction of particle production on the space-time leads to a change of the coherent state $\ket{N}$ into a different $\ket{N'}$, which is either a different coherent state 
 or a decohered one. On the order of one 
 background graviton leaves the coherent state  each Hubble time, due to decay into free quanta or due to rescattering.
     
  After a macroscopic number of gravitons $\Delta N$ of the order of $N$ has decayed,  the resulting  quantum state $\ket{N - \Delta N}$ can no longer -- even approximately -- reproduce the initial de Sitter metric and the classical description stops being valid. Consequently, we obtain the quantum break-time
\begin{align}
t_{\text{q}} \sim  \Gamma^{-1} N \sim  \frac{N}{\sqrt{\Lambda}} \,, \label{eqn:Sca:QuantumBreakTime}
\end{align}
in agreement with \cite{Dvali:2013Compositness, Dvali:2014Quantum}.
Rewriting in terms of more conventional parameters, we get
\begin{align}
t_{\text{q}} \sim {M_P^2 \over \Lambda^{{3\over 2}}} \,. \label{BT}
\end{align}
 When we compare this to the classical break-time \eqref{eqn:Cla:classicalBreakTime} and take into account that the gravitational coupling is $\alpha = 1/N$, we get the same relation \eqref{2times} between the classical and quantum break-times as for the anharmonic oscillator:
\begin{align}
	t_{\text{q}} = \frac{t_{\text{cl}}}{\alpha} \,.
\end{align}
 As there, we argue that $t_{\text{q}}$ is physically meaningful even though $t_{\text{q}} > t_{\text{cl}}$. The reason for this is that $t_{\text{cl}}$ could be increased by a better choice of operators $\hat{a}^{\dagger}$, $\hat{a}$, which take into account classical nonlinear interactions. But for any choice of operators, the decoherence mechanism will continue to work and lead to a significant deviation from the classical solution after $t_{\text{q}}$. Thus, \eqref{BT} is the quantum break-time of de Sitter. 

\subsubsection*{Emergent nature of de Sitter symmetry and breaking 
thereof}

 We would like to briefly comment on symmetry properties of the $1/N$-effects which lead to a quantum break-time of de Sitter. 
 Obviously, since these effects cause departure from the semi-classical evolution, they are obliged not to respect the de Sitter invariance.  
 
  The above fact is fully consistent with our quantum approach. 
  In the standard treatment of gravitational backgrounds, space-times with
  different values of the cosmological constant are considered as different vacua. The corresponding symmetries of such classical backgrounds are thus viewed as {\it vacuum symmetries}. 
   The novelty of our picture is to treat the de Sitter state associated to a certain value of the cosmological constant as a particular quantum state constructed in a Fock space with a unique fundamental vacuum. 
 We have chosen this fundamental vacuum state to be Minkowski. 
 For us de Sitter is therefore a sort of "excited" multi-particle state on top of the Minkowski vacuum. 
 
 Crudely put, instead of being a label of a theory, the cosmological constant in our approach becomes a label of a subset of states in a theory with a unique fundamental vacuum. At this point, the only way we can justify this splitting of roles between Minkowski and de Sitter is by a proof of existence within a simplified model: We managed to construct a model which consistently describes such a situation within its domain of validity.  
 
  Of course, once we "demote" de Sitter from the rank of a vacuum 
 into an ordinary coherent state, its symmetry acquires the meaning  of an emergent symmetry: It is not a  symmetry of the vacuum, but a symmetry of an expectation value over a particular state. In such a situation, an arbitrary process which affects the expectation value is expected to violate this emergent symmetry. This is exactly what is achieved by $1/N$-effects which change the de Sitter coherent state and lead to a finite quantum break-time.

\subsubsection*{Relationship to black holes?}

 It is interesting to note that the quantum break-time of de Sitter space 
\eqref{BT} is reminiscent of the black hole half-life time if we identify 
the de Sitter Hubble radius with the black hole Schwarzschild   
radius.  While for a black hole the importance of this time-scale has a very clear interpretation, for de Sitter it is an absolutely novel feature. 

Moreover, whereas in the black hole case one could at least qualitatively  
envisage the expected state of the system after elapsing of this time in form of a black hole with half of its original mass, for de Sitter the analogous state is completely an open question.  Note that it would be a complete mistake to interpret the resulting state as a de Sitter space-time with a modified value of the cosmological constant.  Rather, we should expect 
that after $t_\text{q}$, de Sitter evolves into an intrinsically-quantum state with no description in form of a solution of classical general relativity. 

 For black holes the story is somewhat different.  
 Since they are localized objects, we should distinguish their exterior and interior regions. 
 Even after the half-life time, the  exterior asymptotic region of a black hole  should still be well-describable by a classical Schwarzschild geometry.   
 
 However, their near-horizon and interior descriptions could become 
 intrinsically quantum. In fact, within the black hole quantum portrait \cite{Nico, Dvali:2011NPortrait, Dvali:2012:PhaseTransition, Dvali:20142ToN, other},  
there are clear indications that just like for 
de Sitter, the $1/N$-corrections to black hole dynamics 
should lead to a breakdown of the classical description of the black hole interior.
Understanding of the precise meaning of this breakdown requires a further 
study.   A very general lesson derived from \cite{Nico} is that the 
quantum break-time is shorter for multi-particle states which are close to criticality (i.e., $\alpha N =1$) and which are characterized by semi-classical Lyapunov exponents.  

The next logical step would be to generalize the analysis of the present work
to the black hole interior by finding an appropriate model for its
quantum description which in the mean field reproduces a classical metric
for the initial times. This would allow for an explicit computation of the quantum break-time along the lines presented here.   
 
\subsection{Bound on number of particle species} 
It is known \cite{SpeciesBound} that semi-classical black hole physics puts a strict upper  bound on the number of particle species ${\mathcal N_{\text{sp}}}$
in terms of the gravity cutoff scale $L_*$:  
\begin{equation}
 {\mathcal N_{\text{sp}}} < {L_*^2 \over L_P^2} \, .
\label{Lbound}
\end{equation} 
That is, the fundamental cutoff length of gravity, $L_*$,  in the presence of  species is no longer given by the Planck 
length $L_P$, but becomes larger and is given by  
$L_* = L_P \sqrt{\mathcal N_{\text{sp}}}$. 
 This bound originates from the fact that the rate of Hawking radiation is proportional to ${\mathcal N_{\text{sp}}}$. Consequently, the evaporation of black holes of size smaller than $L_P \sqrt{\mathcal N_{\text{sp}}}$  cannot be thermal -- even approximately -- due to a very strong back reaction from the decay. Thus, black holes beyond  this size are in conflict with basic properties of Hawking radiation and cannot be treated semi-classically. That is, the scale $L_*$ marks the boundary of applicability of semi-classical 
 Einstein gravity.  Hence, the bound \eqref{Lbound}.  
 
   It is interesting to study the effect of the number of species on the de Sitter quantum break-time. So, let us assume in our simple model that the graviton is coupled to a large number of particle species $\hat{\Psi}_j, ~~j = 1,2,\ldots,{\mathcal N_{\text{sp}}}$.  The presence of more species opens up more channels for Gibbons-Hawking particle production so that the rate increases by a factor of ${\mathcal N_{\text{sp}}}$. Correspondingly, the quantum break-times becomes shorter so that equation (\ref{BT}) takes the form:  
 \begin{align}
t_{\text{q}} \sim {1 \over \sqrt{\Lambda}} \left ({1 \over  {\mathcal N_{\text{sp}}} }  {M_P^2 \over  \Lambda }  \right ) \,. \label{BTSP}
\end{align}
 This relation reveals a very interesting new meaning of the black hole bound on species \eqref{Lbound} in the context of de Sitter. Namely, when species exceed the critical number, 
 \begin{equation}
 {\mathcal N_{\text{cr}}} \equiv {M_P^2 \over \Lambda}, 
 \label{Ncrit}
 \end{equation}
 the quantum break-time becomes shorter than the Hubble time ${1 \over \sqrt{\Lambda}}$. At the same time, the de Sitter radius $R_H$ becomes shorter than the gravity cutoff length $L_*$. 
 
  Moreover, notice that ${\mathcal N_{\text{cr}}} = N$. Thus, the maximal number of particle species allowed in any theory which can provide a de Sitter metric as a trustable classical solution cannot exceed the mean occupation number of de Sitter coherent state gravitons. If we violate this bound, everything goes wrong: The quantum break-time of de Sitter becomes shorter than the Hubble time and the de Sitter radius becomes shorter than the quantum gravity cutoff. 

  We see that, first, there is a nice consistency between the bounds derived from  very different considerations: Two seemingly different 
   requirements -- namely that the de Sitter radius on the one hand and the quantum break-time on the other hand should not exceed the gravity cutoff -- lead to the same conclusion.    
  Secondly, the corpuscular picture of de Sitter gives a very transparent meaning to the species bound: With too many particle species available, the constituent gravitons would decay so rapidly that de Sitter  would  not  last even for a time equal to its radius. Of course, this would make no sense.  The theory protects itself from entering such a nonsense regime, which  would require the 
 {\it classical} de Sitter space to have a curvature radius shorter than the quantum gravity length $L_*$, which is impossible. 
 
 \section{Implications for the cosmological constant} 
 \label{sec:CosmologicalConstant}
 
  As suggested in \cite{Dvali:2014Quantum}, the concept of quantum break-time effectively promotes the cosmological constant problem from an issue of naturalness into a question of consistency. 
  For any given value of the cosmological constant, the quantum break-time puts a consistency constraint on the classically describable duration of the universe. The older it is and the more species it contains, the
 lower is the bound on cosmological constant. Any patch with a given  value of the Hubble parameter $H = \sqrt{\Lambda}$ can be described classically  
  at most during the time $t_{\text{q}} \sim (H^{-1} / {\mathcal N_{\text{sp}}}) (M_P^2/H^2) $, as stated in equation \eqref{BTSP}. 
 
 It is interesting to apply this constraint to our Universe where the observed value of the cosmological constant is $\sqrt{\Lambda} = 10^{-42}\,$GeV. Let us see how long it is possible to describe it classically. Currently, the phenomenologically acceptable number of hidden sector species is  bounded by ${\mathcal N_{\text{sp}}} \sim 10^{32}$ because a larger number of species would lower the gravity cutoff below the TeV-scale, which is excluded by current 
 collider data \cite{SpeciesBound}.  Assuming this number, the observed value of the cosmological term would saturate the quantum break-time bound
 if our Universe  were approximately $10^{100}$ years old. This age should not be confused with the Hubble time or the Hubble radius. It pertains to the entire duration of the classically describable history of our patch.   
 
 We can also apply the constraint in the other direction: Knowing the classical age of a given universe and the number of species, we can deduce an upper bound on $\Lambda$. For the present age $t_{\text{u}}$ of our Universe, we obtain $H_{\text{max}} \sim (M_p^2/ t_{\text{u}})^{1/3} = 10^{-1}\,$GeV, where we set $\mathcal{N}_{\text{sp}} = 1$ to obtain a more robust bound. This means that the energy density associated to the cosmological constant, $\rho \propto H^2 Mp^2$, can be at most $10^{-40}$ of the Planckian energy density.\footnote
 {
    Including $N_{\text{sp}} \sim 10^{32}$ would lead to $H_{\text{max}} \sim  10^{-12}\,$GeV and an energy density of at most $10^{-62}$ of the Planckian value.
 }
  On the one hand, we cannot explain why the cosmological constant is as small as it is. On the other hand, however, the cosmological constant has to be small for consistency.

	\section{Conclusion}	
	\label{sec:Conclusion}
 In this paper, we have first introduced the concepts of classical and quantum break-times and derived some very general relations between them. 
The classical break-time is the time-scale after which classical nonlinearities fully change the time evolution described by a free system. In contrast, the quantum break-time is the time-scale after which the system can no longer be studied classically, no matter how well one accounts for classical nonlinearities. 
   
When the classical state is resolved quantum-mechanically as a coherent state,  systems can generically be understood in terms of two independent expansion parameters. These are the inverse occupation number of quanta, $1/N$, and a four-point quantum coupling, $\alpha$. Both of these parameters vanish in the classical limit, but the quantity which remains finite is $\alpha N$. It can be regarded as the collective coupling in a given state.  On very general grounds, we have shown that $\alpha$ determines the relation between the classical and quantum break-times: $t_{\text{q}} = \, t_{\text{cl}} / \alpha$ (see equation \eqref{2times}). The origin of this relation is clear: 
In the weak field situation, classical nonlinearities correspond to corrections which are suppressed by powers of 
$\alpha N$. These lead to the classical break-time of a linear theory. In contrast, the quantum break-time is due to effects of corpuscular resolution, which are suppressed by powers of $1/N$. Those lead to the quantum break-time of the system. Comparing these two time-scales, we get agreement with \eqref{2times}. Thus, we describe a generic phenomenon: Any classical solution which can be understood as a coherent state of some non-zero frequency quanta must acquire a finite quantum break-time as soon as the quantum interactions among the constituents are taken into account.  
 
 In the main part of the present paper, following \cite{Dvali:2013Compositness}, we
 have first attempted to give a quantum-corpuscular description of de Sitter space in terms of a coherent state and then to apply the concept of quantum and classical break-times to it. In order to make these ideas well-defined and explicit, we have come  up with a simplified but fully consistent quantum model of  spin-2, which has the following key properties: \\
  1)  Classically, it provides a solution which fully reproduces the de Sitter metric for short enough time-scales. \\
 2)  This classical de Sitter space-time can be represented as the expectation value of a corresponding quantum graviton field over a well-defined coherent state $\ket{N}$. \\
 We have shown that  de Sitter coherent state $\ket{N}$ corresponds to a
  multi-graviton state of frequencies given by the de Sitter Hubble parameter $\sqrt{\Lambda}$ and mean occupation number $N = M_P^2 / \Lambda$. 
   
    Next, we took into account the coupling of the coherent state gravitons to external probe particles $\hat{\Psi}$.   When the graviton field is treated classically, the quantum evolution of $\hat{\Psi}$-particles reproduces all the usual features of fields quantized on top of a fixed classical de Sitter background metric. However, in the microscopic language in which de Sitter is treated as a quantum state, we see very clearly  which quantum processes are responsible for generating the known semi-classical effects and which ones are  overlooked by the classical treatment. In particular,  we have shown that the quantum decay and rescattering of coherent state gravitons lead to the production of $\hat{\Psi}$-particles.  In the appropriate limit, this phenomenon matches Gibbons-Hawking particle production in de Sitter space. It is remarkable that one can reproduce the well known semi-classical properties of de Sitter space within a simple and fully-quantum theory in which de Sitter is just another quantum state. 
  
   However, the main novelty in our quantum framework is that  we are able to capture back reaction effects, which are invisible in the semi-classical treatment.  Namely, we have shown that Gibbons-Hawking particle production and other quantum processes necessarily change an initial quantum coherent state. This back reaction inevitably leads to decoherence and as a result to a quantum break-time of de Sitter space. This quantum break-time is given by \eqref{BT}.

 By no means do we pretend to have understood the complete quantum 
 picture of de Sitter space in full nonlinear Einstein gravity. 
 However, the model we work with, despite being simple, 
 takes us surprisingly far in reproducing the known properties of de Sitter 
 space in a fundamentally new quantum language.  Of course, this may
 very well be simply a remarkable coincidence, but can also be an indication 
 that we are on the right track in understanding the quantum nature of de Sitter space.   
     
   Needless to say, the existence of quantum corpuscular effects which lead to a quantum break-time for seemingly-eternal classical spaces, such as de Sitter, can have important cosmological implications. Some of them were already discussed in \cite{Dvali:2013Compositness} and \cite{Dvali:2014Quantum}, where the idea of a corpuscular resolution of de Sitter (as well as anti-de Sitter) space was put forward.  We shall mention a couple of obvious open questions.\\  Can 
 the existence of a finite quantum break time provide a natural graceful exist for Guth's original inflationary scenario \cite{Guth}? As it is well known, in this scenario  inflation takes place in a meta-stable 
  quasi-de Sitter state, with exponentially-long lifetime.  In contrast, the quantum-break time of de Sitter space discussed in this paper is only power-law long.  The natural question to ask would be whether by taking it into account, the graceful exit in Guth's scenario  can take place without any need of tunneling?\\
  The second obvious question concerns the fate of the cosmological constant. What does it evolve to in the presence of a finite quantum break-time?  Whatever the answer is, at least such questions allow us to put the long standing problem of the cosmological constant in a new light. As already pointed out in \cite{Dvali:2014Quantum}, the existence of a de Sitter quantum break-time may promote the cosmological constant problem from an issue of naturalness into a question of quantum consistency. 
  The quantum break-time tells us that the older the classical universe is and the more particle species it houses, the lower is the upper bound on its vacuum energy.\footnote
  {
  	The age of the Universe is counted throughout the classical history of a given patch and should not be confused with the Hubble 
  radius. Otherwise, the statement would be trivial.
  }
  For example, assuming the maximal number of phenomenologically acceptable species, the observed value of the cosmological constant would saturate the consistency bound if our Universe were $10^{100}$ years old.  
  
 In the same way, the quantum break-time puts a consistency constraint on
 the maximal duration of the inflationary stage with given Hubble parameter.  The  classical description of an inflationary universe with Hubble parameter $H$ and number of species  ${\mathcal N_{\text{sp}}}$ 
 can only be trusted for a number of e-foldings not exceeding~${1 \over {\mathcal N_{\text{sp}}}} M_P^2/H^2$.      
  
 A natural question within the corpuscular approach to the cosmological constant would be how to extend the analysis  to anti-de Sitter space. As mentioned above, the picture of anti-de Sitter as multi-particle quantum state was already put forward in earlier work \cite{Dvali:2013Compositness}, where it was noticed that the natural mean occupation number of constituent quanta $N$ coincides with a central charge of the conformal field theory dictated by the AdS/CFT correspondence \cite{AdsCFT}. We leave the application of our approach to anti-de Sitter space for future work \cite{AdSF}.

   It would be interesting to understand whether there exists any connection 
  between our concept of quantum break-time of de Sitter and other possible instabilities, such as, for example, the one claimed in \cite{polyakov}. At first sight, no obvious connection is visible since our effect crucially relies on the quantum coherent state resolution of the de Sitter metric, which is a fundamentally new ingredient.  

   As a very final remark, the universality of the relations which classical and quantum break-times obey in seemingly different theories (spin-2 versus anharmonic oscillations of scalar field) suggests that concepts developed in this paper can find universal applications. For instance, anharmonic oscillations of the scalar field can be viewed as a model  describing coherent oscillations of the axion field in the early universe and therefore enable us to study the influence of quantum physics on axion evolution. From our results it follows that this influence is negligible. 
   
   \section*{Acknowledgements}
   The work of G.D. was supported by the Humboldt Foundation under Alexander von Humboldt Professorship, the ERC Advanced Grant  "Selfcompletion"  (Grant No. 339169), FPA 2009-07908, CPAN (CSD2007-00042), HEPHACOSP-ESP00346, and by TR 33 "The Dark Universe". 
   The work of C.G. was supported in part by the Humboldt Foundation and by Grants: FPA 2009-07908, CPAN (CSD2007-00042), and by the ERC Advanced Grant 339169 "Selfcompletion".

	\appendix
	
	\section{Calculation of the rate of particle production}
	\label{sec:AppendixParticleProduction}
	
In order to obtain the correct prefactors for the normalizations which we use, we rederive how a general $S$-matrix element determines the differential decay rate. We consider an initial coherent state $\ket{N}$, in which a constituent quantum of energy $m$ decays to two particles with 4-momenta $p_1=(p_{0,1}, \vec{p}_1)$ and $p_2=(p_{0, 2}, \vec{p}_2)$. This leads to the final state $\ket{N'} \otimes \ket{\text{f}_\Psi}$, where $\ket{N'}$ is a possibly different coherent state and $\ket{\text{f}_\Psi} = \hat{b}_{\vec{p}_1}^\dagger \hat{b}_{\vec{p}_2}^\dagger \ket{0}$ describes the two external particles. The differential transition probability is given by the square of the $S$-matrix element $\mathcal{A}$, divided by the norms of final and initial state and multiplied by the phase space factor:
	\begin{align*}
	\diff \text{w}_{\text{fi}} = \frac{\left|\mathcal{A}\right|^2}{\braket{N|N}\braket{N'|N'}\braket{\text{f}_\Psi|\text{f}_\Psi}}  \frac{\diff^3 \vec{p}_1\, V}{(2\pi)^3 } \frac{\diff^3 \vec{p}_2\, V}{(2\pi)^3 } = \left|\mathcal{A}\right|^2 \diff^3 \vec{p}_1 \diff^3 \vec{p}_2\,, 
	\end{align*}
where we used that coherent states are normalized, $\braket{N|N} = \braket{N'|N'} =1$, and that $\braket{\text{f}_\Psi|\text{f}_\Psi} = \left(\delta^{(3)}(\vec{0})\right)^2 = \left(V/(2\pi)^3\right)^2$. Defining the Feynman amplitude $\mathcal{M}$ via
	\begin{align*}
\mathcal{A} = (2\pi)^4 \delta(m-p_{0,1} - p_{0,2})\delta^{(3)}(\vec{p}_1 - \vec{p}_2) \mathcal{M} \,, 
	\end{align*}
we obtain the differential rate
	\begin{align*}
	\diff \Gamma  = \frac{\diff \text{w}_{\text{fi}}}{T}= |\mathcal{M}|^2 (2\pi)^4 V \delta(m-p_{0,1} - p_{0,2}) |\vec{p}_1|^2\, \diff |\vec{p}_1| \, \diff^2 \Omega \,,
	\end{align*} 
		where we regularized the divergence of the one-dimensional $\delta$-distribution with the help of the time $T$ during which the reaction happens, $\delta(0) = \frac{T}{2\pi}$. Evaluating the last $\delta$-distribution, we get
	\begin{align*}
	\frac{\diff\Gamma}{\diff \Omega} = \frac{|\vec{p}_1|V |\mathcal{M}|^2}{16\pi^2 m \zeta_\Psi(p_1)^2 \zeta_\Psi(p_2)^2 }  \,, 
	\end{align*}
	with $\zeta_\Psi(p) = \left( (2\pi)^3 2 p_0\right)^{-1/2}$.
	
For our application to particle production, the $S$-matrix element \eqref{eqn:Sca:ProductionSMatrix} yields the Feynman amplitude
	\begin{align*}
	\mathcal{M} = \frac{1}{\sqrt{2m}}  \mathcal{K}(-p,p') \sqrt{\frac{N'}{V}}\left(1-\frac{\Delta N^2}{8N}\right) \,. 
	\end{align*}
	Thus, we get for the differential decay constant:
	\begin{align*}
	\frac{d\Gamma}{d\Omega} = \frac{ \sqrt{\frac{m^2}{4} -m_\Psi^2} N'}{2\pi M_p^2 m^2  }  \left( p_1 \cdot p_2  + 2 m_{\Psi}^2 \right)^2 \left(1-\frac{\Delta N^2}{4N}\right)  \,,
	\end{align*}
	where we plugged in \eqref{eqn:Sca:KinematicalFactorRedshift}. After integrating over the angles, we obtain the final decay constant to leading order in $1/N$:
\begin{align*}
	\Gamma = \frac{2 \sqrt{\frac{m^2}{4} -m_\Psi^2} N}{ M_p^2 m^2 }\left( p_1 \cdot p_2 + 2 m_{\Psi}^2 \right)^2 \left(1-\frac{\Delta N^2 - 4\Delta N}{4N}\right) \,.
	\end{align*}

\end{document}